\documentclass[aps,prc,floatfix,nofootinbib,showpacs,superscriptaddress,twocolumn]{revtex4-1}

\usepackage{amsmath}
\usepackage{amssymb}
\usepackage{amsthm}
\usepackage{dcolumn}
\usepackage{epsfig}
\usepackage{graphics}
\usepackage{graphicx}
\usepackage{amsmath}
\usepackage{longtable}
\usepackage{color}

\definecolor{purple}{rgb}{0.5,0,0.5}
\definecolor{blue}{rgb}{0.0,0,0.9}

\begin{document}
\title{Nucleon and Roper electromagnetic elastic and transition form factors}

\author{D.\,J.~Wilson}
\affiliation{Physics Division, Argonne National Laboratory, Argonne, Illinois 60439, USA}

\author{I.\,C.~Clo\"et}
\affiliation{CSSM and CoEPP, School of Chemistry and Physics
University of Adelaide, Adelaide SA 5005, Australia}

\author{L.~Chang}
\affiliation{Physics Division, Argonne National Laboratory, Argonne, Illinois 60439, USA}

\author{C.\,D.~Roberts}
\affiliation{Physics Division, Argonne National Laboratory, Argonne, Illinois 60439, USA}
\affiliation{Institut f\"ur Kernphysik, Forschungszentrum J\"ulich, D-52425 J\"ulich, Germany}
\affiliation{Department of Physics, Illinois Institute of Technology, Chicago, Illinois 60616, USA}

\begin{abstract}
We compute nucleon and Roper electromagnetic elastic and transition form factors using a Poincar\'e-covariant, symmetry-preserving treatment of a vector$\,\times\,$vector contact-interaction.  Obtained thereby, the electromagnetic interactions of baryons are typically described by hard form factors.  In contrasting this behaviour with that produced by a momentum-dependent interaction, one achieves comparisons which highlight that elastic scattering and resonance electroproduction experiments probe the evolution of the strong interaction's running masses and coupling to infrared momenta.  For example, the existence, and location if so, of a zero in the ratio of nucleon Sachs form factors are strongly influenced by the running of the dressed-quark mass.
In our description of the nucleon and its first excited state, diquark correlations are important.  These composite and fully-interacting correlations are instrumental in producing a zero in the Dirac form factor of the proton's $d$-quark; and in determining the ratio of $d$-to-$u$ valence-quark distributions at $x=1$, as we show via a simple formula that expresses $d_v/u_v(x=1)$ in terms of the nucleon's diquark content.
The contact interaction produces a first excitation of the nucleon that is constituted predominantly from axial-vector diquark correlations.  This impacts greatly on the $\gamma^\ast p \to P_{11}(1440)$ form factors, our results for which are qualitatively in agreement with the trend of available data.  Notably, our dressed-quark core contribution to $F_{2\ast}(Q^2)$ exhibits a zero at $Q^2 \approx 0.5\,m_N^2$.
Faddeev equation treatments of a hadron's dressed-quark core usually underestimate its magnetic properties, hence we consider the effect produced by a dressed-quark anomalous electromagnetic moment.  Its inclusion much improves agreement with experiment.
On the domain $0 < Q^2\lesssim 2\,$GeV$^2$, meson-cloud effects are conjectured to be important in making a realistic comparison between experiment and hadron structure calculations.  We find that our computed helicity amplitudes are similar to the bare amplitudes inferred via coupled-channels analyses of the electroproduction process.  This supports a view that extant hadron structure calculations, which typically omit meson-cloud effects, should directly be compared with the bare-masses, -couplings, etc., determined via coupled-channels analyses.
\end{abstract}


\maketitle


\section{Introduction}
Building a bridge between QCD and the observed properties of hadrons is one of the key problems in modern science.  The international programme focused on the physics of excited nucleons is close to the heart of this effort.  It addresses the questions: which hadron states and resonances are produced by QCD, and how are they constituted?  The $N^\ast$ program therefore stands alongside the search for hybrid and exotic mesons as an integral part of the search for an understanding of QCD.  An example of the theory activity in this area is provided in Ref.\,\cite{Aznauryan:2009da}.

It is in this context that we consider the $N(1440)P_{11}$, $J^P=(1/2)^+$ Roper resonance, whose discovery was reported in 1964 \cite{Roper:1964zz}.  In important respects the Roper appears to be a copy of the proton.  However, its (Breit-Wigner) mass is 50\% greater \cite{Nakamura:2010zzi}.  This feature has long presented a problem within the context of constituent-quark models formulated in terms of colour-spin potentials, which typically produce the following level ordering \cite{Capstick:2000qj}: ground state, $J^P=(1/2)^+$ with radial quantum number $n=0$ and angular momentum $l=0$; first excited state, $J^P=(1/2)^-$ with $(n,l)=(0,1)$; second excited state, $J^P=(1/2)^+$, with $(n,l)=(1,0)$; etc.  The difficulty is that the lightest $l=1$ baryon appears to be the $N(1535)S_{11}$, which is heavier than the Roper.  Holographic models of QCD, viewed by some as a covariant generalisation of constituent-quark potential models, predict degeneracy of the $(n,l)=(1,0)$ and $(0,1)$ states \cite{deTeramond:2005su}.  Whilst it has been observed that constituent-quark models with Goldstone-boson exchange potentials can produce the observed level ordering \cite{Glozman:1997ag}, such a foundation makes problematic a unified description of baryons and mesons.

In order to correct the level ordering problem within the potential model paradigm, other ideas have been explored.  The possibility that the Roper is simply a hybrid baryon with constituent-gluon content is difficult to support because the lightest such states occur with masses above $1.8\,$GeV \cite{Capstick:1999qq}.  An alternative is to consider the presence of explicit constituent-$\bar q q$ components within baryon bound-states \cite{JuliaDiaz:2006av}.  Whilst not literally correct,
such a picture may be interpreted as suggesting that $\pi N$ final-state interactions must play an important role in any understanding of the Roper.  This perspective is common to modern coupled-channels treatments of baryon resonances \cite{Gasparyan:2003fp,Suzuki:2009nj,Doring:2010ap}, and finds support in contemporary numerical simulations of lattice-QCD \cite{Mahbub:2010rm} and Dyson-Schwinger equation (DSE) studies \cite{Roberts:2011cf,Roberts:2011rr,Roberts:2011ym}.

Given that an understanding of the Roper has long eluded practitioners, it is unsurprising that this resonance has been a focus of the $N^\ast$ programme
at Jefferson Lab (JLab).  Experiments at JLab \cite{Aznauryan:2008pe,Dugger:2009pn,Aznauryan:2009mx,Aznauryan:2011td} have enabled an extraction of nucleon-to-Roper transition form factors and thereby exposed the first zero-crossing seen in any nucleon form factor or transition amplitude.  Explaining this new structure also presents a challenge for theory \cite{Aznauryan:2011qj}.

Notwithstanding its history, an understanding of the Roper is perhaps now beginning to emerge through a constructive interplay between dynamical coupled-channels models and hadron structure calculations, particularly those symmetry-preserving studies made using the tower of Dyson-Schwinger equations \cite{Maris:2003vk,Roberts:2007jh,Holt:2010vj,Chang:2011vu}.  One indication of this is found in predictions for the masses of the baryons' dressed-quark-cores \cite{Roberts:2011cf}, which match the bare masses of nucleon resonances determined by the Excited Baryon Analysis Center (EBAC) \cite{Suzuki:2009nj} with a rms-relative error of 14\% and, in particular, agree with EBAC's value for the bare-mass of the Roper resonance; viz. (in GeV),
\begin{equation}
\label{Ropermass}
m_{\rm Roper}^{QQQ}=1.82\pm 0.07\;\; \mbox{cf.} \;\;
m_{\rm Roper}^{\rm EBAC-bare} = 1.76 \pm 0.10\,.
\end{equation}
The DSE state is the first excitation of the ground-state nucleon whilst the EBAC bare state is the source for three distinct features in the $\pi N$-scattering $P_{11}$ partial wave, which migrate widely from the real-energy axis once meson-nucleon final-state interactions are enabled. It is notable that the dressed-quark core of the nucleon's parity partner is approximately 400\,MeV heavier than $m_{\rm Roper}^{QQQ}$ and 1.1\,GeV heavier than the core of the ground-state nucleon, a magnitude commensurate with its origin in dynamical chiral symmetry breaking (DCSB) \cite{Roberts:2011cf}.

\begin{figure}[t]
\centerline{%
\includegraphics[clip,width=0.45\textwidth]{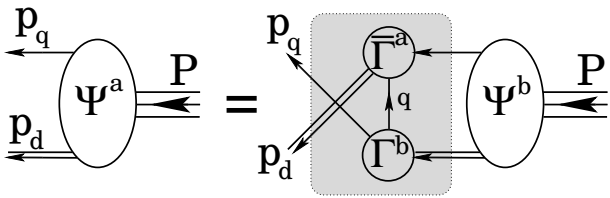}}
\caption{\label{fig:Faddeev} Poincar\'e covariant Faddeev equation, Eq.\,(\protect\ref{FEone}), employed herein to calculate baryon properties.  $\Psi$ in Eq.\,(\protect\ref{PsiNucleon}) is the Faddeev amplitude for a baryon of total momentum $P= p_q + p_d$.  It expresses the relative momentum correlation between the dressed-quark and -diquarks within the baryon.  The shaded region demarcates the kernel of the Faddeev equation, Sec.\,\protect\ref{sec:Faddeev}, in which: the \emph{single line} denotes the dressed-quark propagator, Sec.\,\protect\ref{sec:gap};
$\Gamma$ is the diquark Bethe-Salpeter amplitude, Sec.\,\protect\ref{qqBSA};
and the \emph{double line} is the diquark propagator, Eqs.\,(\ref{scalarqqprop}), (\ref{avqqprop}).}
\end{figure}

Herein we probe further into the possibility that $\pi N$ final-state interactions play a critical role in understanding of the Roper, through a simultaneous computation within the DSE framework of nucleon and Roper elastic form factors, and the form factors describing the nucleon-to-Roper transition.  In so doing we add materially to a body of work that presents the unified analysis of many properties of meson and baryon ground- and excited-states based on the symmetry-preserving treatment of a single quark-quark interaction; namely, a vector-vector contact-interaction.  This procedure has already been applied to the spectrum of $u,d$-quark mesons and baryons \cite{Roberts:2011cf}, and the electromagnetic properties of $\pi$- and $\rho$-mesons, and their diquark partners \cite{GutierrezGuerrero:2010md,Roberts:2010rn,Roberts:2011wy}.  These studies provide the foundation for much of that which follows.

In Sec.\,\ref{framework} we present a brief overview of our framework: both the Faddeev equation treatment of the nucleon and Roper dressed-quark cores, and the currents which describe the interaction of a photon with a baryon composed from consistently-dressed constituents.  Additional material is expressed in appendices and referred to as necessary.
In Sec.\,\ref{sec:nucleonelastic} we describe the parameter-free calculation of nucleon elastic form factors within a DSE treatment of the contact interaction.  Germane to our presentation are comparisons both with data and computations using QCD-like momentum-dependence for the propagators and vertices.  In addition, we use the elastic form factors to predict the ratio of valence-quark distribution functions at $x=1$.

We begin to describe our results for the Roper elastic and nucleon-to-Roper transition form factors in Sec.\,\ref{sec:transitionelastic}.  The description continues in Sec.\,\ref{sec:HAMM}, with a consideration of the impact on all form factors of a dressed-quark anomalous magnetic moment.  In Sec.\,\ref{sec:mesoncloud} we explore the effect of meson-cloud contributions to hadron structure calculations in the context of the $\gamma^\ast p \to P_{11}(1440)$ helicity amplitudes, which have been analysed using coupled-channels methods \cite{Matsuyama:2006rp,Suzuki:2010yn,JuliaDiaz:2009ww,LeePrivate:2011}.

Section~\ref{sec:summary} is an epilogue.

\begin{table}[t]
\caption{(A)~Computed quantities required as input for the Faddeev equation, obtained with $\alpha_{\rm IR}/\pi =0.93$ and (in GeV) $m=0.007$, $\Lambda_{\rm ir} = 0.24\,$, $\Lambda_{\rm uv}=0.905$.
(B)~Nucleon and Roper masses, and associated unit-normalised eigenvectors, obtained therewith.
(All dimensioned quantities are listed in GeV.)
\label{Table:FE}
}
\begin{center}
\begin{tabular*}
{\hsize}
{
c@{\extracolsep{0ptplus1fil}}
c@{\extracolsep{0ptplus1fil}}
c@{\extracolsep{0ptplus1fil}}
|c@{\extracolsep{0ptplus1fil}}
c@{\extracolsep{0ptplus1fil}}
c@{\extracolsep{0ptplus1fil}}
c@{\extracolsep{0ptplus1fil}}}\hline
$M$ & $m_{qq 0^+}$ & $m_{qq 1^+}$ & $E_{qq 0^+}$ & $F_{qq 0^+}$ & $E_{qq 1^+}$ & $M d^{1/2}_{\cal F}$~ \\
0.368 & 0.776 & 1.056 & 4.354 & 0.499 & 1.3029 & 0.880~
\\\hline
\end{tabular*}
\end{center}
\begin{center}
\begin{tabular*}
{\hsize}
{
c@{\extracolsep{0ptplus1fil}}
|c@{\extracolsep{0ptplus1fil}}
c@{\extracolsep{0ptplus1fil}}
c@{\extracolsep{0ptplus1fil}}
c@{\extracolsep{0ptplus1fil}}
c@{\extracolsep{0ptplus1fil}}}\hline
\mbox{mass~(GeV)} & $s$ & $a_1^+$ & $a_1^0$ & $a_2^+$ & $a_2^0$ \\\hline
$m_N= 1.14$ & ~~0.88 & -0.38~ & ~0.27 & -0.065& 0.046\\
$m_R= 1.72$ & -0.44 & -0.030 & ~\;0.021 & ~0.73 & -0.52~~~\\\hline
\end{tabular*}
\end{center}
\end{table}

\section{Electromagnetic Currents}
\label{framework}
We base our description of the dressed-quark-core of the nucleon and Roper on solutions of a Faddeev equation, which is illustrated in Fig.\,\ref{fig:Faddeev}, and formulated and described in Apps.\,\ref{sec:contact}, \ref{sec:Faddeev}.  The Faddeev equations are completed by the quantities reported in Table~\ref{Table:FE}A, and our values for the nucleon and Roper masses and eigenvectors, the latter normalised to unity, are presented in Table~\ref{Table:FE}B.
These masses are drawn from a unified spectrum of $u,d$-quark hadrons, obtained using a symmetry-preserving regularisation of a vector$\,\times\,$vector contact interaction \cite{Roberts:2011cf}.  That study simultaneously correlates the masses of meson and baryon ground- and excited-states within a single framework.  In comparison with relevant quantities, it produces a root-mean-square-relative-error$/$degree-of-freedom equal to 13\%.
The predictions uniformly overestimate the experimental values of meson and baryon masses \cite{Nakamura:2010zzi}.  Given that the employed truncation deliberately omitted meson-cloud effects in the Faddeev kernel, this is a good outcome because inclusion of such contributions acts to reduce the computed masses.  As noted in the Introduction, Eq.\,\eqref{Ropermass}, such effects are particularly important for the Roper resonance.

\begin{figure}[t]
\begin{centering}
\includegraphics[clip,width=0.30\textwidth]{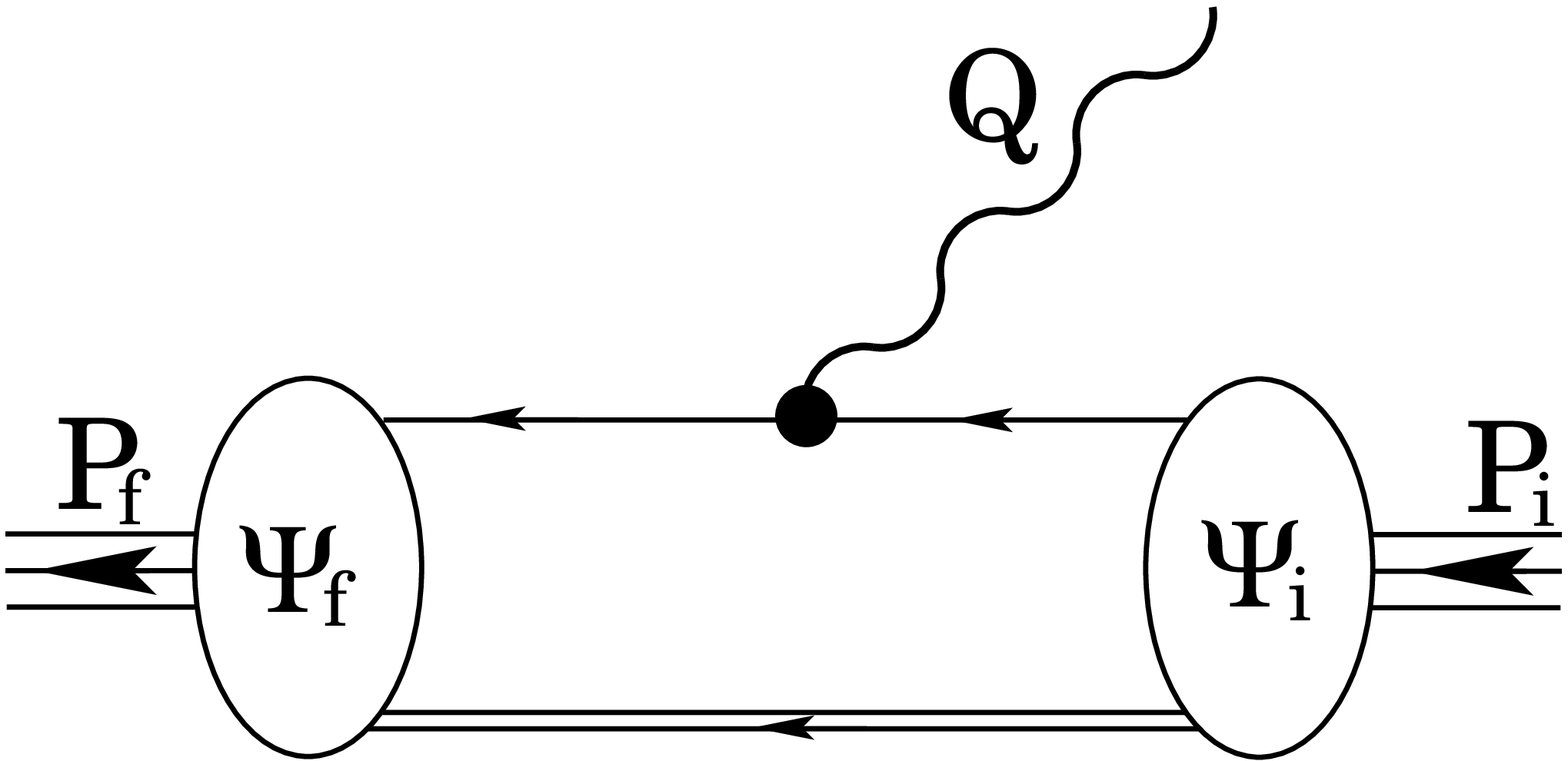}
\vspace*{1em}

\includegraphics[clip,width=0.30\textwidth]{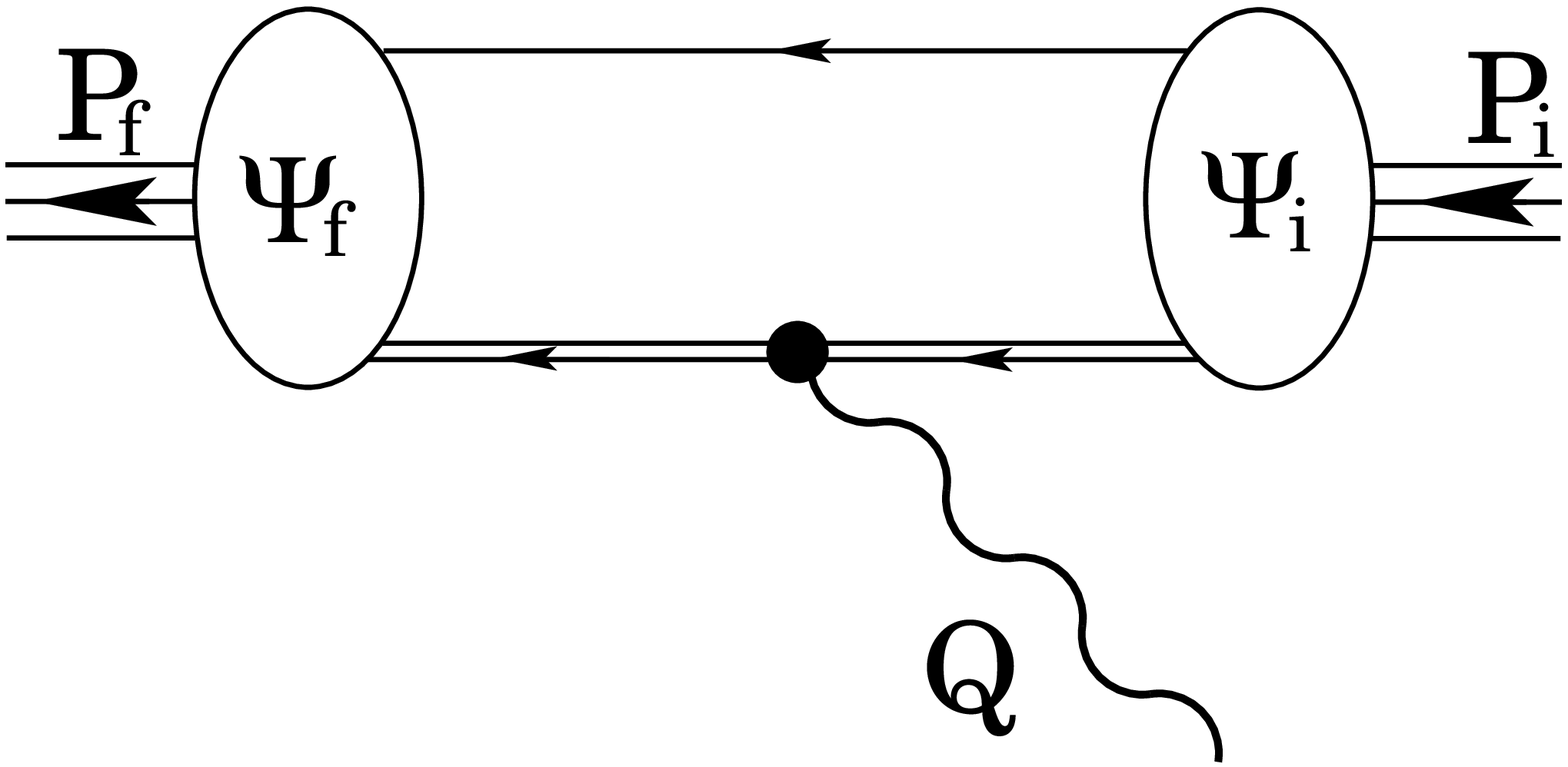}
\vspace*{1em}

\includegraphics[clip,width=0.30\textwidth]{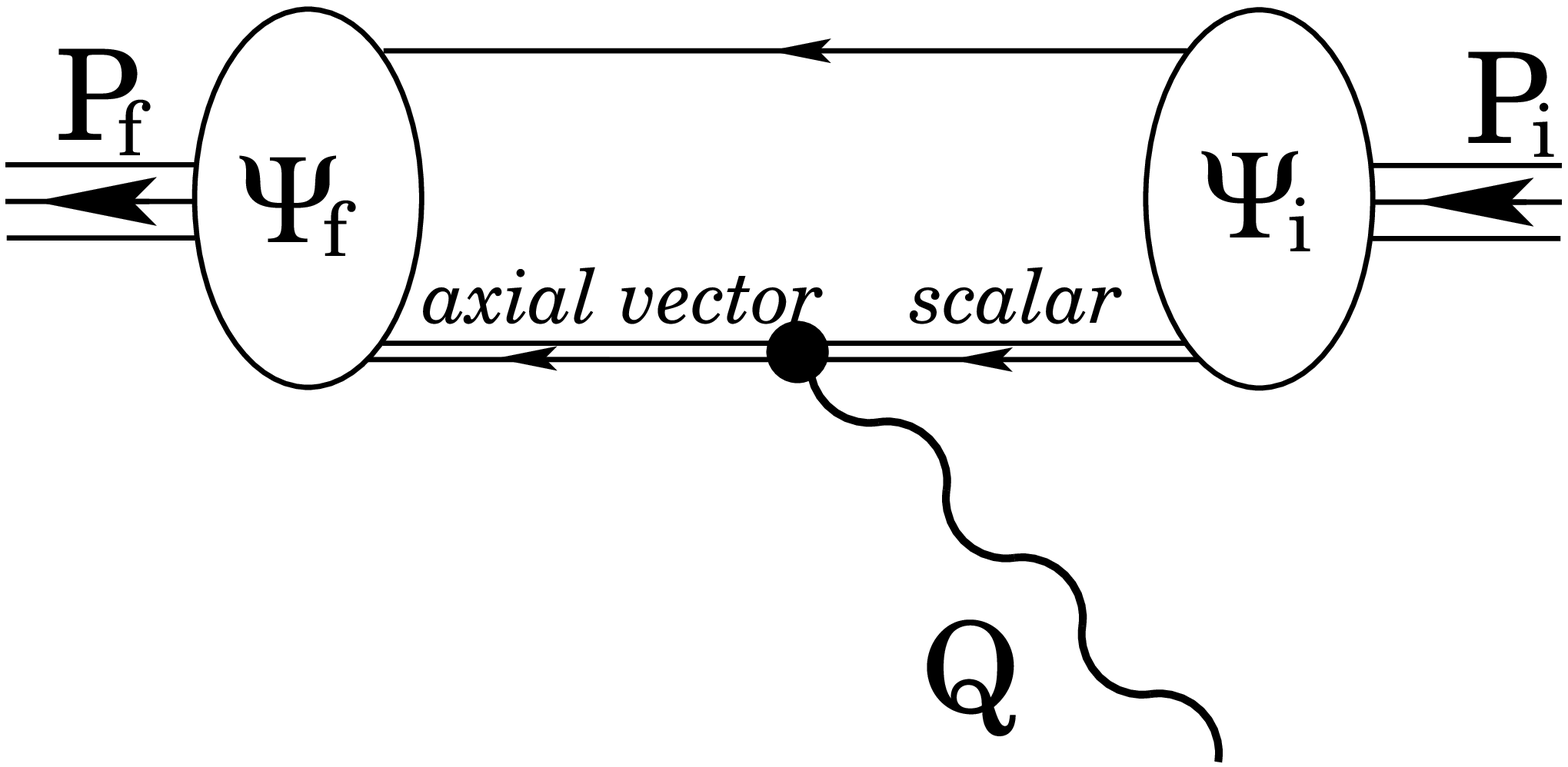}
\end{centering}
\caption{\label{fig:current} Interaction vertex which ensures a conserved current for the elastic and transition form factors in Eqs.\,(\protect\ref{Belastic}), (\protect\ref{NRtransition}).  The single line represents the dressed-quark propagator, $S(p)$ in App.\,\protect\ref{sec:gap}; the double line, the diquark propagators in Eqs.\,(\protect\ref{scalarqqprop}) and (\protect\ref{avqqprop}); and the vertices are described in App.\,\protect\ref{App:current}.  From top to bottom, the diagrams describe the photon coupling: directly to the dressed-quark; to a diquark, in an elastic scattering event; or inducing a transition between scalar and axial-vector diquarks.}
\end{figure}

We are interested in three electromagnetic currents: those defining the nucleon and Roper elastic  form factors
\begin{eqnarray}
\nonumber
J_\mu^B(P_f,P_i) & = & i e\,  \bar u_{B}(P_f)\left[ \gamma_\mu F_{1B}(Q^2) \right. \\
&& \left. + \frac{1}{2 M_{B}} \sigma_{\mu\nu} Q_\nu F_{2B}(Q^2)\right] u_{B}(P_i),
\label{Belastic}
\end{eqnarray}
$B=N$, $R$ and $Q=P_f - P_i$; and that expressing the transition form factors [$Q_\mu \gamma_\mu^T = 0$, Eq.\,(\ref{GammaQ})]
\begin{eqnarray}
\nonumber
J_\mu^\ast(P_f,P_i) & = & i e\,  \bar u_{R}(P_f)\left[ \gamma_\mu^T F_{1\ast}(Q^2) \right. \\
&& \left. + \frac{1}{M_R+M_{N}} \sigma_{\mu\nu} Q_\nu F_{2\ast}(Q^2)\right] u_{N}(P_i).\rule{1em}{0ex}
\label{NRtransition}
\end{eqnarray}
N.B.\ Electromagnetic current kinematics and the definition of constraint-independent form factors are discussed in Ref.\,\cite{Devenish:1975jd}, so that Eq.\,(\ref{Belastic}) may be viewed as a special case of Eq.\,(\ref{NRtransition}) which is simplified by the on-shell condition $\bar u_B(P_f) \gamma\cdot Q u_B(p_i) = 0$.

With the contact interaction described in App.\,\ref{sec:contact} and our treatment of the Faddeev equation, App.\,\ref{sec:Faddeev}, there are three contributions to the currents.  They are illustrated in Fig.\,\ref{fig:current} and detailed in App.\,\ref{App:current}.  The computation of form factors is straightforward following the procedures outlined in those appendices.

\begin{table}[t]
\caption{Row~1: Results computed herein with the contact interaction, whose input is presented in Table~\ref{Table:FE}.  Row~2: Results obtained using QCD-like momentum-dependence for the dressed-quark propagators and diquark Bethe-Salpeter amplitudes in solving the Faddeev equation.  Row~3: Values representative of experiment.  Row~4: Contact interaction augmented by a model dressed-quark anomalous electromagnetic moment (see Sec.\,\protect\ref{sec:HAMM}).
\label{tab:nucleonstatic}
}
\begin{center}
\begin{tabular*}
{\hsize}
{
l@{\extracolsep{0ptplus1fil}}
c@{\extracolsep{0ptplus1fil}}
c@{\extracolsep{0ptplus1fil}}
c@{\extracolsep{0ptplus1fil}}
c@{\extracolsep{0ptplus1fil}}
c@{\extracolsep{0ptplus1fil}}
c@{\extracolsep{0ptplus1fil}}}
    & $r_{1p} M_N$ & $r_{2p} M_N$ & $r_{1n} M_N$ & $r_{2n} M_N$ & $\kappa_p$& $\kappa_n$\\\hline
contact & 3.19 & 2.84 & 1.21 & 3.19 & 1.02 & -0.92 \\
Ref.\,\protect\cite{Cloet:2008re} & 3.76 & 2.82 & 0.59 & 3.14 & 1.67 & -1.59 \\
Ref.\,\protect\cite{Kelly:2004hm} & 3.76 & 4.18 & 0.56 & 4.33 & 1.79 & -1.91 \\
contact$_{\rm QAMM}$ & 3.41 & 4.00 & 0.55 & 3.85 & 1.68 & -1.24 \\\hline
\end{tabular*}
\end{center}
\end{table}

\section{Nucleon Elastic}
\label{sec:nucleonelastic}
\begin{figure}[t]
\begin{centering}
\includegraphics[clip,width=0.45\textwidth]{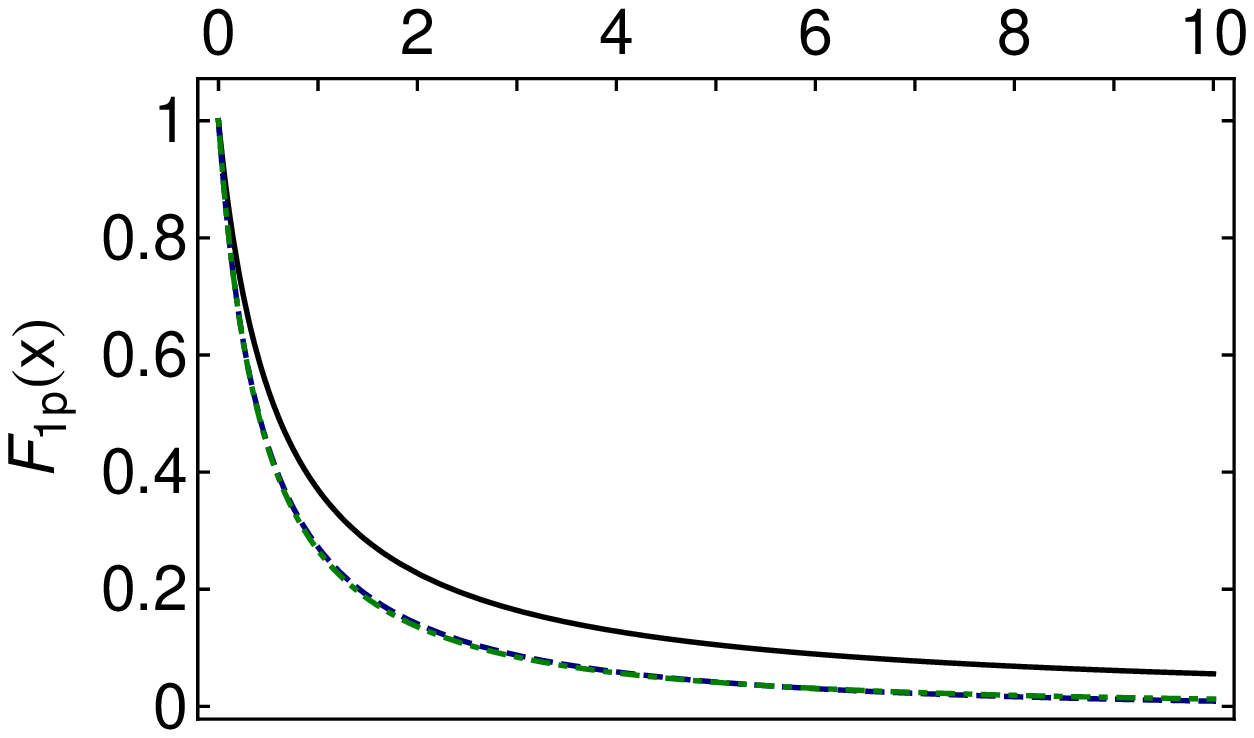}
\vspace*{-4.5ex}

\includegraphics[clip,width=0.45\textwidth]{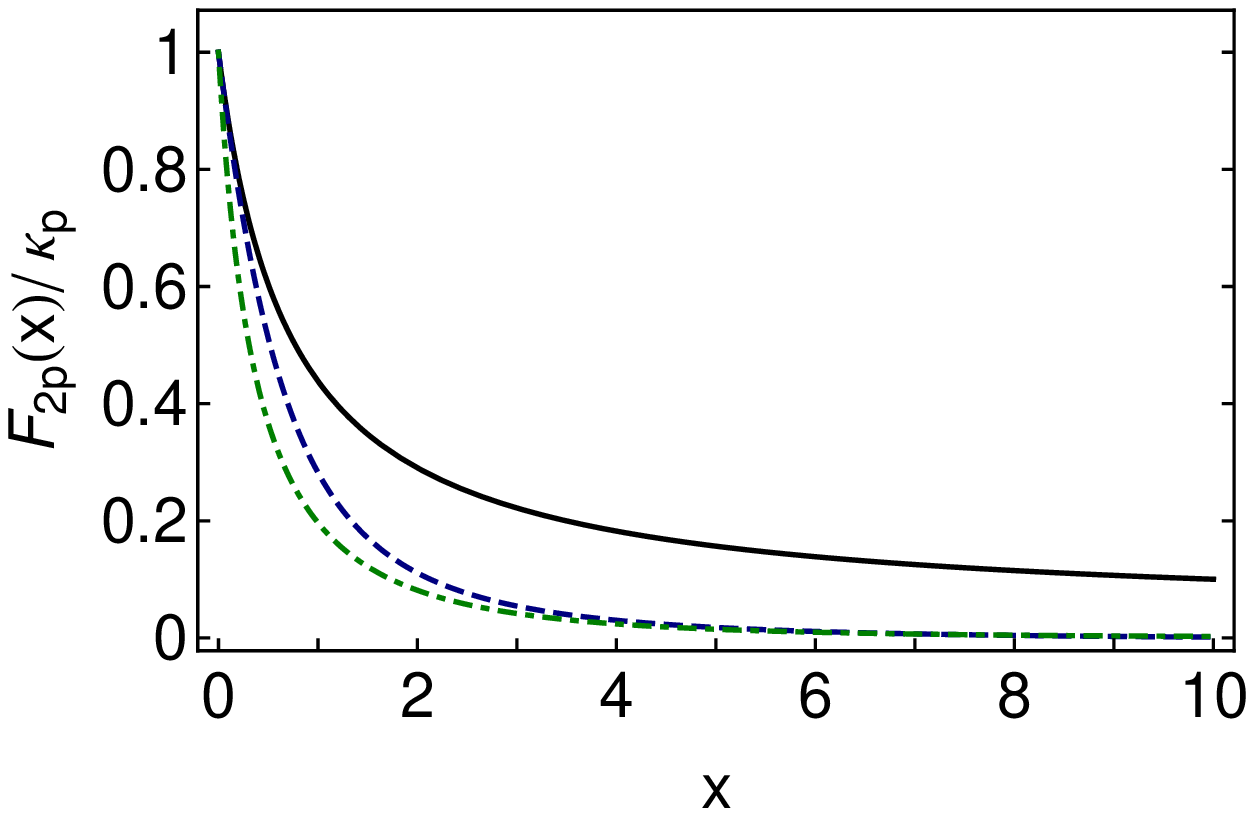}
\end{centering}
\caption{\label{fig:ProtonF1F2} Proton Dirac (upper panel) and Pauli (lower panel) form factors, as a function of $x=Q^2/m_N^2$.  \emph{Solid curve} -- result obtained herein using the contact-interaction and hence a dressed-quark mass-function and diquark Bethe-Salpeter amplitudes that are momentum-independent; \emph{dashed curve} -- result obtained in Ref.\,\protect\cite{Cloet:2008re}, which employed QCD-like momentum-dependence for the dressed-quark propagators and diquark Bethe-Salpeter amplitudes in solving the Faddeev equation; \emph{dot-dashed curve} -- a parametrisation of experimental data \protect\cite{Kelly:2004hm}.}
\end{figure}

There are no free parameters in our computation of nucleon elastic form factors: all those associated with our treatment of the contact interaction are fixed in Refs.\,\cite{Roberts:2011wy,Roberts:2011cf}, see Table~\ref{Table:FE}.  We report static properties in Table~\ref{tab:nucleonstatic}, and depict form factors for the proton in Fig.\,\ref{fig:ProtonF1F2} and the neutron in Fig.\,\ref{fig:NeutronF1F2}.  N.B.\ We use a Euclidean metric, App.\,\ref{App:EM}, and hence in elastic scattering one has
\begin{equation}
P_f^2 = - m_B^2 = P_i^2\,,\; Q^2 + 2 P_i\cdot Q = 0\,,
\end{equation}
where $m_B$ is the mass of the baryon involved.

\subsection{Dirac and Pauli Form factors}
\begin{figure}[t]
\begin{centering}
\includegraphics[clip,width=0.45\textwidth]{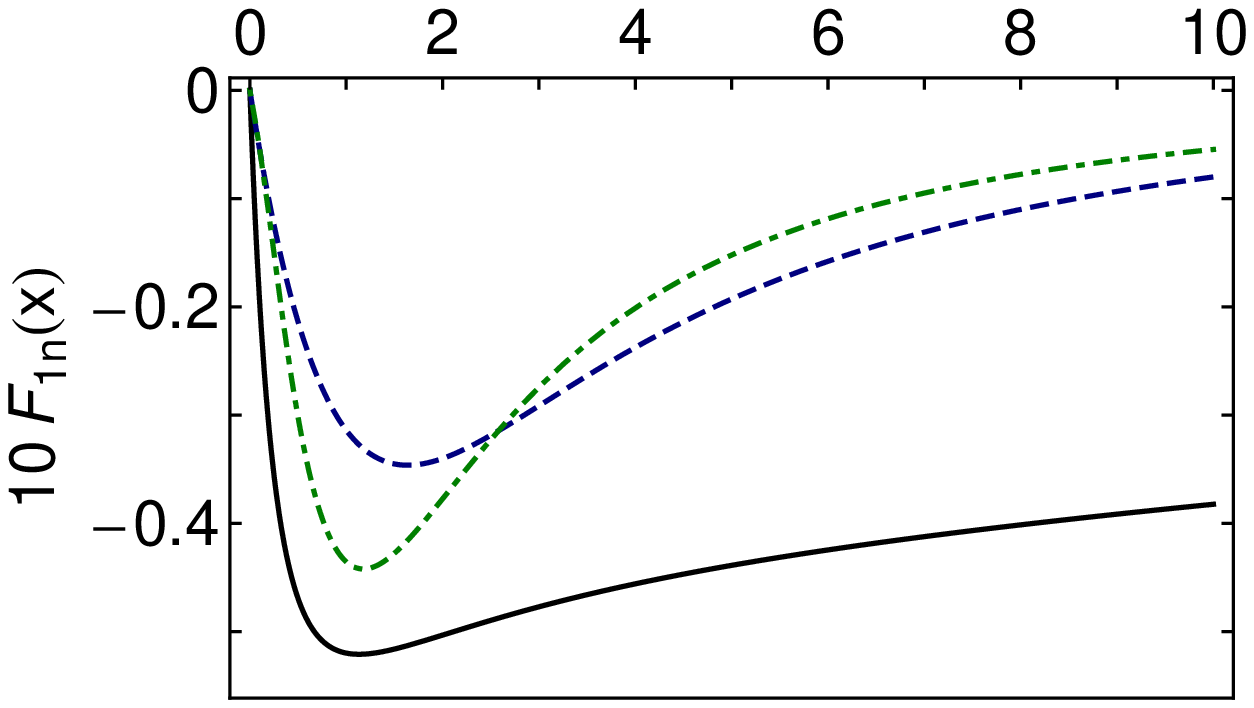}
\vspace*{-4.5ex}

\includegraphics[clip,width=0.45\textwidth]{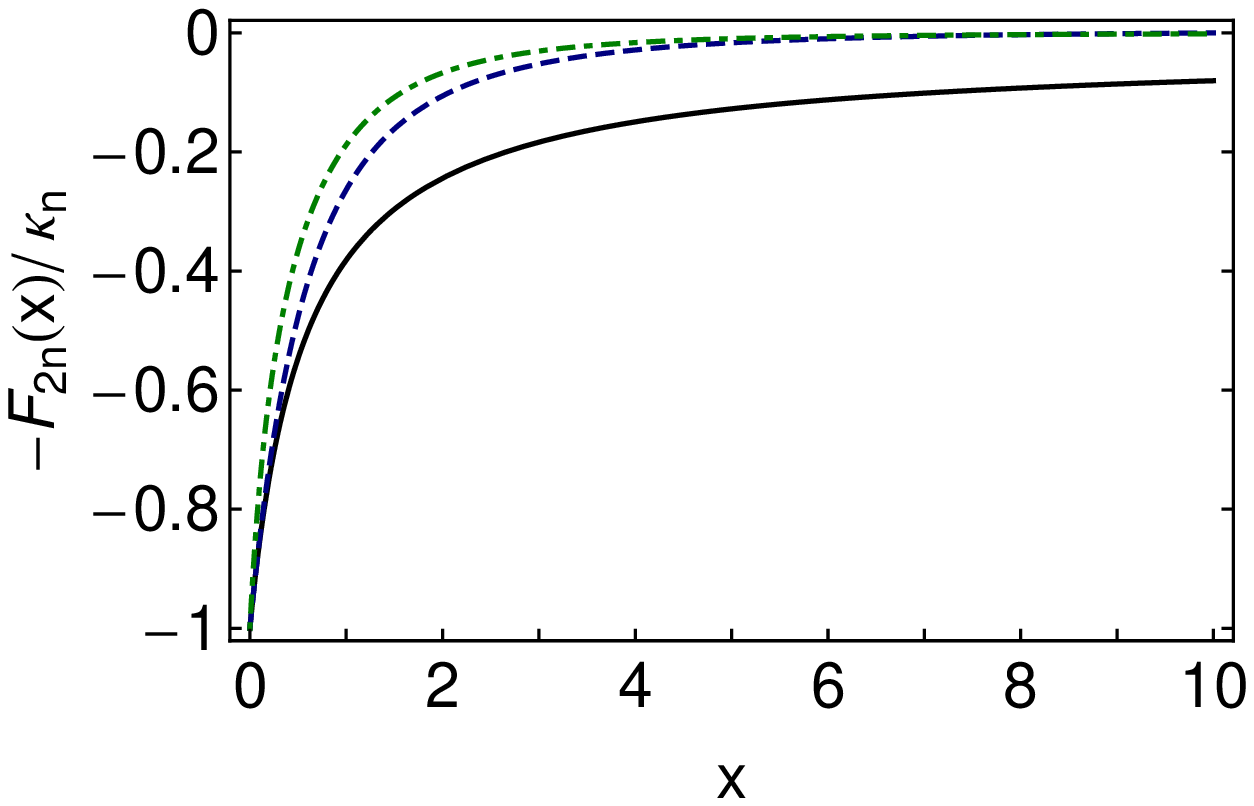}
\end{centering}
\caption{\label{fig:NeutronF1F2} Neutron Dirac (upper panel) and Pauli (lower panel) form factors, as a function of $x=Q^2/m_N^2$.  \emph{Solid curve} -- result obtained herein using the contact-interaction and hence a dressed-quark mass-function and diquark Bethe-Salpeter amplitudes that are momentum-independent; \emph{dashed curve} -- result obtained in Ref.\,\protect\cite{Cloet:2008re}, which employed QCD-like momentum-dependence for the dressed-quark propagators and diquark Bethe-Salpeter amplitudes in solving the Faddeev equation; \emph{dot-dashed curve} -- a parametrisation of experimental data \protect\cite{Kelly:2004hm}.}
\end{figure}

In our symmetry-preserving DSE-treatment of the contact interaction we construct a nucleon from diquarks whose Bethe-Salpeter amplitudes are momentum-independent and dressed-quarks with a momentum-independent mass-function, and arrive at a nucleon described by a momentum-independent Faddeev amplitude.  This last is the hallmark of a pointlike composite particle and explains the hardness of the computed form factors, which is evident in Figs.\,\ref{fig:ProtonF1F2}, \ref{fig:NeutronF1F2}.

The hardness contrasts starkly with results obtained from a momentum-dependent Faddeev amplitude produced by dressed-quark propagators and diquark Bethe-Salpeter amplitudes with QCD-like momentum-dependence; and with experiment.  Evidence for a connection between the momentum-dependence of each of these elements and the behaviour of QCD's $\beta$-function is accumulating; e.g., Refs.\,\cite{Maris:2000sk,Bhagwat:2006pu,GutierrezGuerrero:2010md,Roberts:2010rn,Roberts:2011wy,%
Eichmann:2008ef,Eichmann:2011vu}.
The comparisons in Figs.\,\ref{fig:ProtonF1F2}, \ref{fig:NeutronF1F2} add to this evidence, in connection here with readily accessible observables, and support a view that experiment is a sensitive probe of the running of the $\beta$-function to infrared momenta.  This perspective will be reinforced by subsequent figures.

Table~\ref{tab:nucleonstatic} exposes another shortcoming in the description of nucleons via a momentum-independent Faddeev amplitude;
namely, the anomalous magnetic moments are far too small.  In a Poincar\'e-covariant treatment, the magnitude of the magnetic moment grows with increasing quark orbital angular momentum.  However, a momentum-independent Faddeev amplitude suppresses quark orbital angular momentum, as may be seen from the absence in Eqs.\,\eqref{FaddeevAmp} of a dependence on the relative momentum.  This explains the differences between the anomalous magnetic moments in Rows~1 and 2 of Table~\ref{tab:nucleonstatic}.

The differences between the anomalous moments in Rows~2 and 3 have a different origin; viz., QCD's dressed-quarks possess large momentum-dependent anomalous magnetic moments owing to dynamical chiral symmetry breaking \cite{Chang:2010hb}, and the discrepancy is resolved by incorporating this phenomenon.  Owing to the momentum dependence of these moments, the magnetic radii are also affected, so that $r_{2p}$, $r_{2n}$ in Row~2 are shifted markedly toward the values in Row~3.  This is illustrated in Ref.\,\cite{Chang:2011tx} and in Row~4, which is discussed further in Sec.\,\ref{sec:HAMM}.

\begin{figure}[t]
\begin{centering}
\includegraphics[clip,width=0.45\textwidth]{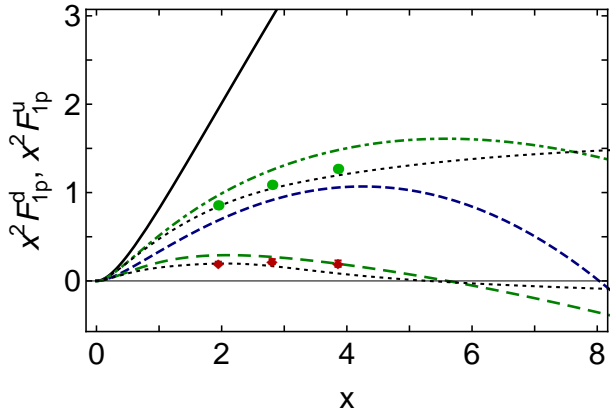}
\end{centering}
\caption{\label{fig:F1uF1d}
Flavour separation of the proton's Dirac form factor, as a function of $x=Q^2/m_N^2$: normalisation: $F_{1p}^u(0)=2$, $F_{1p}^d(0)=1$.
\emph{Solid curve} -- $u$-quark obtained using the contact interaction;
\emph{short-dashed curve} -- $d$-quark, contact interaction;
\emph{dot-dashed curve} -- $u$-quark obtained from QCD-like momentum-dependence for the dressed-quark propagators and diquark Bethe-Salpeter amplitudes in the Faddeev equation \protect\cite{Cloet:2008re}; and
\emph{long-dashed curve} -- $d$-quark obtained similarly.  The data are from Refs.\,\protect\cite{Riordan:2010id,Cates:2011pz}: $u$-quark, circles; and $d$-quark, diamonds.
The \emph{dotted curves} are determined from the parametrisation of data in Ref.\,\protect\cite{Bradford:2006yz}.
}
\end{figure}

In Fig.\,\ref{fig:F1uF1d} we depict a flavour decomposition of the proton's Dirac form factor. In neither the data nor the calculations is the scaling behaviour anticipated from perturbative QCD evident on the momentum domain depicted.  This fact is emphasised by the zero in $F_{1p}^d$, whose existence is independent of the interaction.  Its location is not, and the extrapolation of a modern parametrisation of data produces a zero which is coincident with that predicted by the QCD-based interaction \cite{Cloet:2008re,Cloet:2011qu}.  The zero owes to the presence of diquark correlations in the nucleon.  It has been found \cite{Cloet:2008re} that the proton's singly-represented $d$-quark is more likely to be struck in association with an axial-vector diquark correlation than with a scalar, and form factor contributions involving
an axial-vector diquark are soft.  On the other hand, the doubly-represented $u$-quark
is predominantly linked with harder scalar-diquark contributions.  This interference produces
the zero in the Dirac form factor of the $d$-quark in the proton.  The location of the
zero depends on the relative probability of finding $1^+$ and $0^+$ diquarks in the proton: with increasing probability for an axial-vector diquark, it moves to smaller-$x$ -- in Ref.\,\cite{Cloet:2008re} the scalar-diquark probability is 60\%, whereas herein it is 78\%.

\begin{figure}[t]
\begin{centering}
\includegraphics[clip,width=0.45\textwidth]{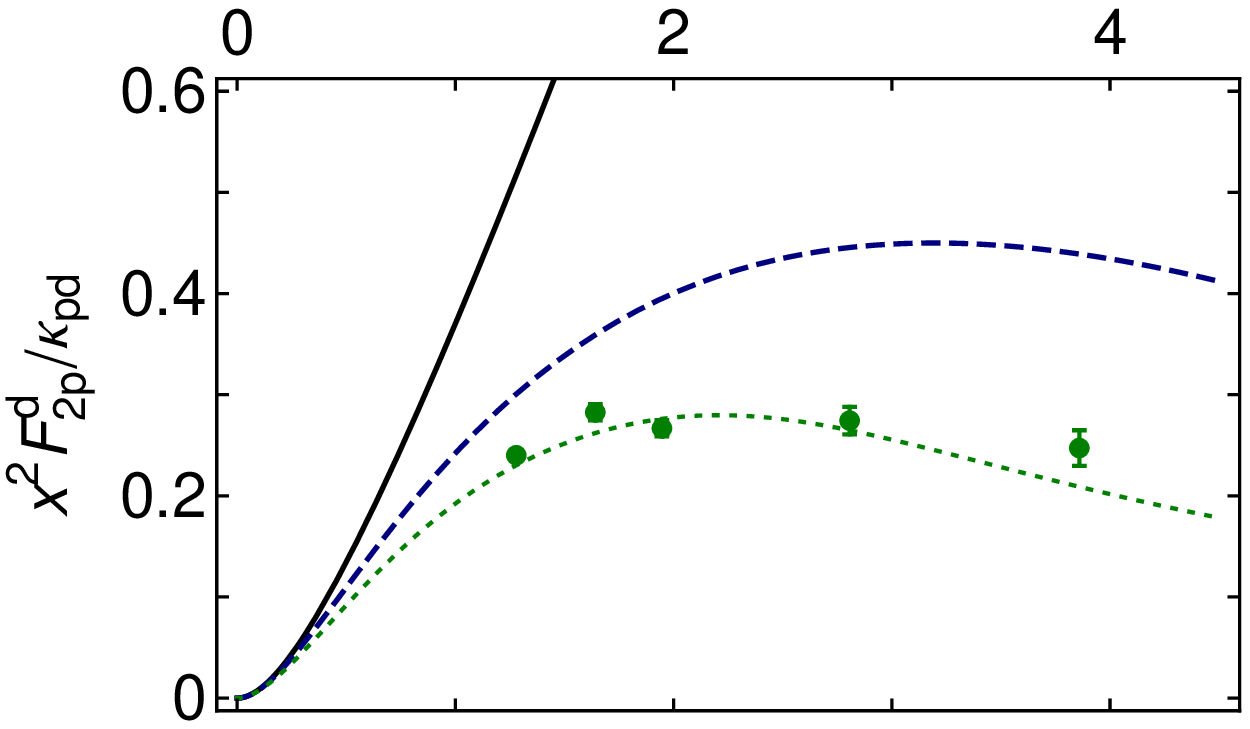}
\vspace*{-11.5ex}

\includegraphics[clip,width=0.45\textwidth]{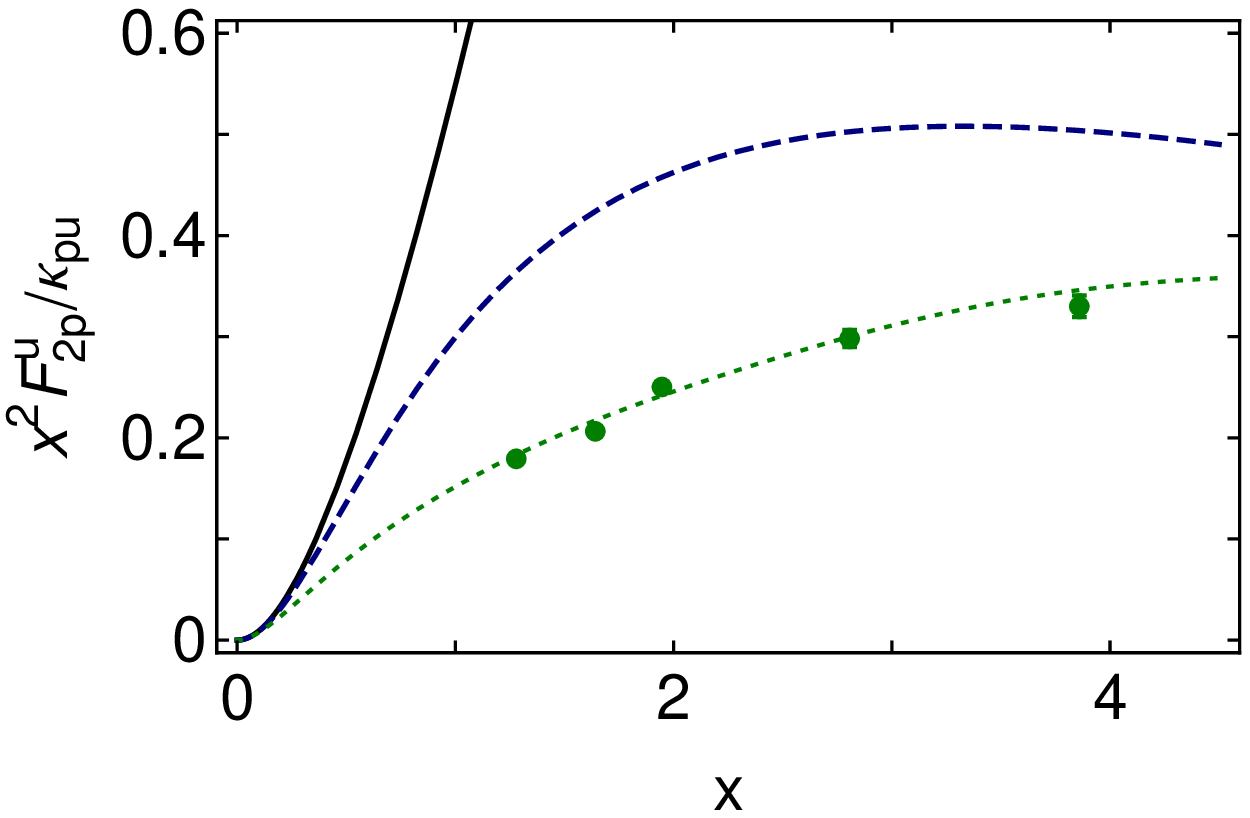}
\end{centering}
\caption{\label{fig:F2dF2u}
Flavour separation of the proton's Pauli form factor, as a function of $x=Q^2/m_N^2$: $d$-quark, upper panel; and $u$-quark, lower panel.
\emph{Solid curve} -- result obtained using the contact interaction; \emph{dashed curve} -- obtained from QCD-like momentum-dependence for the dressed-quark propagators and diquark Bethe-Salpeter amplitudes in the Faddeev equation \protect\cite{Cloet:2008re}; \emph{dotted curve} -- determined from the parametrisation of data in Ref.\,\protect\cite{Bradford:2006yz}; and data from Refs.\,\protect\cite{Plaster:2005cx,Riordan:2010id,Cates:2011pz}.}
\end{figure}

We plot the flavour decomposition of the proton's Pauli form factor in Fig.\,\ref{fig:F2dF2u}.  Once again, the contact-interaction results are far too hard and the general trend of the data favours a Faddeev equation built from dressed-quark propagators and diquark Bethe-Salpeter amplitudes which are QCD-like in their momentum dependence.

\subsection{Sachs form factors}
\begin{figure}[t]
\begin{centering}
\includegraphics[clip,width=0.45\textwidth]{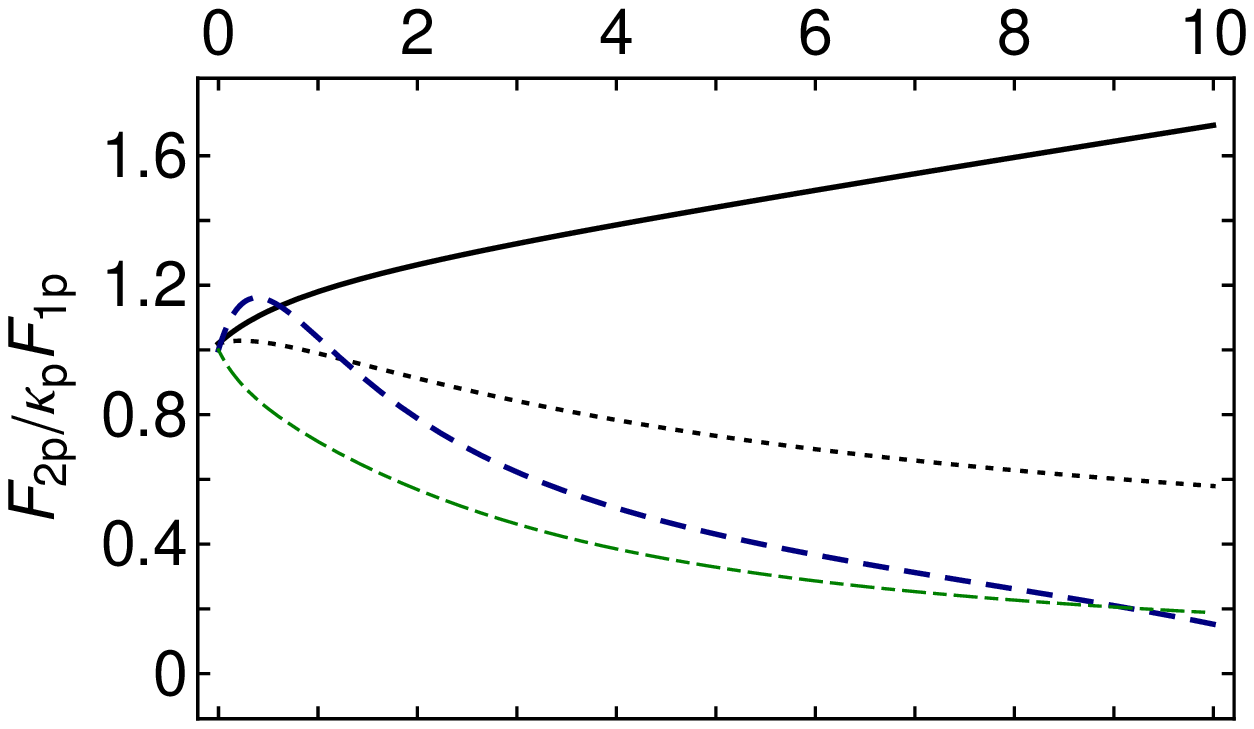}
\vspace*{-11.5ex}

\includegraphics[clip,width=0.45\textwidth]{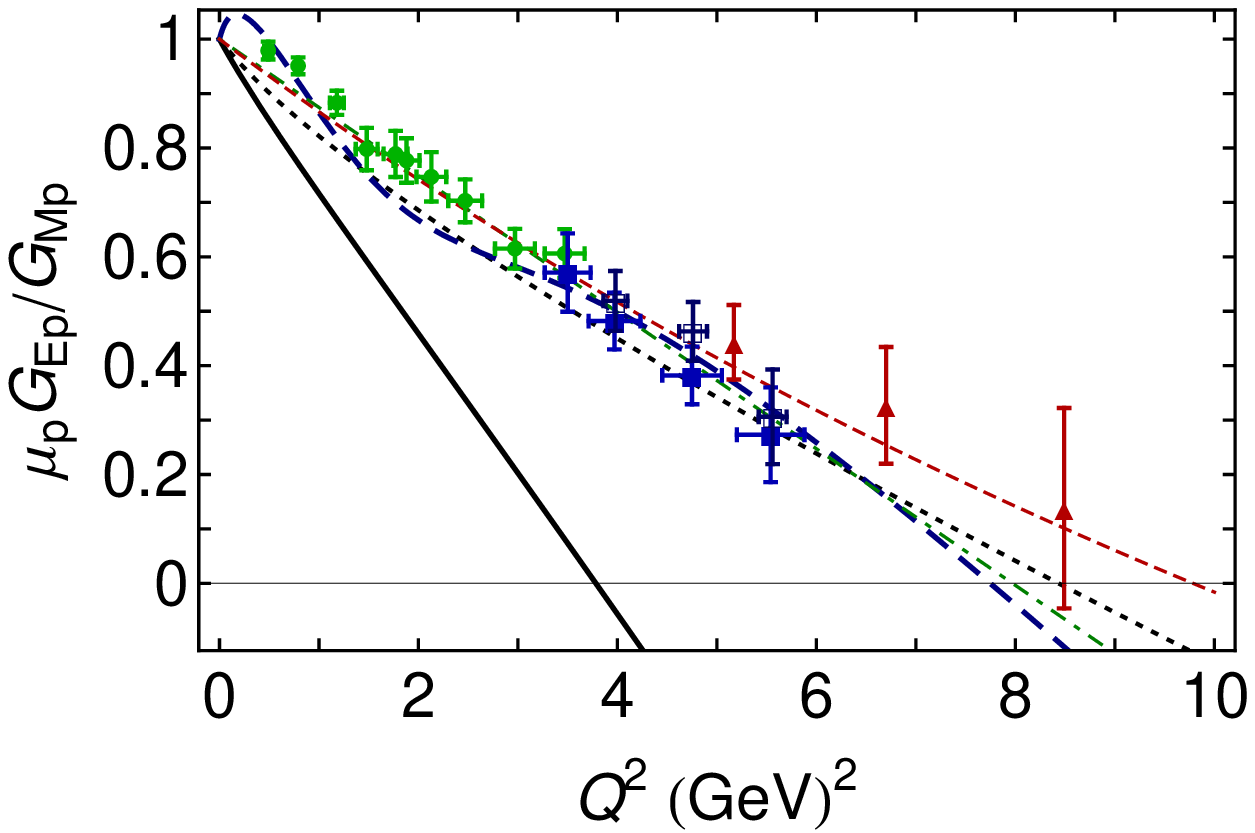}
\end{centering}
\caption{\label{fig:GEGMp}
\emph{Upper panel}: Normalised ratio of proton Pauli and Dirac form factors.  \emph{Solid curve} -- contact interaction; \emph{long-dashed curve} -- result from Ref.\,\protect\cite{Chang:2011tx}, which employed QCD-like momentum-dependence for the dressed-quark propagators and diquark Bethe-Salpeter amplitudes; \emph{long-dash-dotted curve} -- drawn from parametrisation of experimental data in Ref.\,\protect\cite{Kelly:2004hm}; and \emph{dotted curve} -- softened contact-interaction result, described in connection with Eq.\,\eqref{SoftF2p}.
\emph{Lower panel}: Normalised ratio of proton Sachs electric and magnetic form factors.  \emph{Solid curve} and \emph{long-dashed curve}, as above; \emph{dot-dashed curve} -- linear fit to data in Refs.\,\protect\cite{Jones:1999rz,Gayou:2001qd,Qattan:2004ht,Punjabi:2005wq,Puckett:2010ac}, constrained to one at $Q^2=0$; \emph{short-dashed curve} -- $[1,1]$-Pad\'e fit to that data; and \emph{dotted curve} -- softened contact-interaction result, described in connection with Eq.\,\eqref{SoftF2p}.
In addition, we have represented a selection of data explicitly: filled-squares \protect\cite{Gayou:2001qd}; circles \protect\cite{Punjabi:2005wq}; up-triangles \protect\cite{Puckett:2010ac}; and open-squares \protect\cite{Puckett:2011xg}.
}
\end{figure}

The lower panel of Fig.\,\ref{fig:GEGMp} depicts the ratio of proton Sachs electric and magnetic form factors:
\begin{subequations}
\begin{eqnarray}
G_{Ep}(Q^2) &=& F_{1p}(Q^2) - \frac{Q^2}{4 m_N^2} F_{2p}(Q^2),\\
G_{Mp}(Q^2) &=& F_{1p}(Q^2) + F_{2p}(Q^2)\,.
\end{eqnarray}
\end{subequations}
Once again, the existence of a zero is independent of the interaction upon which the Faddeev equation is based but the location is not.  That location is insensitive to the size of the diquark correlations \cite{Cloet:2008re}.

In order to assist in explaining the origin and location of a zero in the Sachs form factor ratio, in the top panel of Fig.\,\ref{fig:GEGMp} we depict the ratio of Pauli and Dirac form factors: both the actual contact-interaction result and that obtained when the Pauli form factor is artificially ``softened;'' viz.,
\begin{equation}
\label{SoftF2p}
F_{2p}(Q^2) \to \frac{F_{2p}(Q^2)}{1+Q^2/(4 m_N^2)}\,.
\end{equation}
As observed in Ref.\,\cite{Bloch:2003vn}, a softening of the proton's Pauli form factor has the effect of shifting the zero to larger values of $Q^2$.  In fact, if $F_{2p}$ becomes soft quickly enough, then the zero disappears completely.

The Pauli form factor is a gauge of the distribution of magnetisation within the proton. Ultimately, this magnetisation is carried by the dressed-quarks and influenced by correlations amongst them, which are expressed in the Faddeev wave-function.
If the dressed-quarks are described by a momentum-independent mass-function, then they behave as Dirac particles with constant Dirac values for their magnetic moments and produce a hard Pauli form factor.
%
Alternatively, suppose that the dressed-quarks possess a momentum-dependent mass-function, which is large at infrared momenta but vanishes as their momentum increases.  At small momenta they will then behave as constituent-like particles with a large magnetic moment, but their mass and magnetic moment will drop toward zero as the probe momentum grows.  (N.B.\ Massless fermions do not possess a measurable magnetic moment \cite{Chang:2010hb}.)  Such dressed-quarks will produce a proton Pauli form factor that is large for $Q^2 \sim 0$ but drops rapidly on the domain of transition between nonperturbative and perturbative QCD, to give a very small result at large-$Q^2$.  The precise form of the $Q^2$-dependence will depend on the evolving nature of the angular momentum correlations between the dressed-quarks.
From this perspective, existence, and location if so, of the zero in $\mu_p G_{Ep}(Q^2)/G_{Mp}(Q^2)$ are a fairly direct measure of the location and width of the transition region between the nonperturbative and perturbative domains of QCD as expressed in the momentum-dependence of the dressed-quark mass-function.

We expect that a mass-function which rapidly becomes partonic -- namely, is very soft -- will not produce a zero; have seen that a constant mass-function produces a zero at a small value of $Q^2$, and know that a mass-function which resembles that obtained in the best available DSE studies \cite{Bhagwat:2006tu,Qin:2011dd} and via lattice-QCD simulations \cite{Bowman:2005vx}, produces a zero at a location that is consistent with extant data.  There is an opportunity here for very constructive feedback between future experiments and theory.

\subsection{Valence-quark distributions at $x=1$}
At this point we would like to exploit a connection between the $Q^2=0$ values of elastic form factors and the Bjorken-$x=1$ values of the dimensionless structure functions of deep inelastic scattering, $F_2^{n,p}(x)$.  Our first remark is that the $x=1$ value of a structure function is invariant under the evolution equations \cite{Holt:2010vj}.  Hence the value of
\begin{equation}
\label{dvuv1}
\left. \frac{d_v(x)}{u_v(x)}\right|_{x\to 1}\rule{-0.5em}{0ex}, \;\mbox{where} \rule{1em}{0ex}
\frac{d_v(x)}{u_v(x)} =
\frac{4 \frac{F_2^n(x)}{F_2^p(x)} - 1}{4- \frac{F_2^n(x)}{F_2^p(x)}},
\end{equation}
is a scale-invariant feature of QCD and a discriminator between models.  Next, when Bjorken-$x$ is unity, then $Q^2+2P\cdot Q=0$; i.e., one is dealing with elastic scattering.  Therefore, in the neighbourhood of $x=1$ the structure functions are determined by the target's elastic form factors.
The ratio in Eq.\,\eqref{dvuv1} expresses the relative probability of finding a $d$-quark carrying all the proton's light-front momentum compared with that of a $u$-quark doing the same or, equally, owing to invariance under evolution, the relative probability that a $Q^2=0$ probe either scatters from a $d$-quark or a $u$-quark; viz.,
\begin{equation}
\label{dvuvF1}
\left. \frac{d_v(x)}{u_v(x)}\right|_{x\to 1} = \frac{P_{1}^{p,d}}{P_{1}^{p,u}}.
\end{equation}
%

Plainly, in $SU(6)$ constituent-quark models, the right-hand-side of Eq.\,\eqref{dvuvF1} is $1/2$.  On the other hand, when a Poincar\'e-covariant Faddeev equation is employed to describe the nucleon,
\begin{equation}
\label{dvuvF1result}
\frac{P_{1}^{p,d}}{P_{1}^{p,u}} =
\frac{\frac{2}{3} P_1^{p,a} + \frac{1}{3} P_1^{p,m}}
{P_1^{p,s}+\frac{1}{3} P_1^{p,a} + \frac{2}{3} P_1^{p,m}},
\end{equation}
where we have used the notation of Ref.\,\cite{Cloet:2008re}.  Namely,
$P_1^{p,s}=F_{1p}^s(Q^2=0)$ is the contribution to the proton's charge arising from diagrams with a scalar diquark component in both the initial and final state: $u[ud]\otimes \gamma \otimes u[ud]$.  The diquark-photon interaction is far softer than the quark-photon interaction and hence this diagram contributes solely to $u_v$ at $x=1$.
$P_1^{p,a}=F_{1p}^a(Q^2=0)$, is the kindred axial-vector diquark contribution; viz., $2 d\{uu\}\otimes \gamma\otimes d\{uu\}+u\{ud\} \otimes\gamma \otimes u\{ud\}$.  At $x=1$ this contributes twice as much to $d_v$ as it does to $u_v$.
$P_1^{p,m}=F_{1p}^m(Q^2=0)$, is the contribution to the proton's charge arising from diagrams with a different diquark component in the initial and final state.  The existence of this contribution relies on the exchange of a quark between the diquark correlations and hence it contributes twice as much to $u_v$ as it does to $d_v$.  If one uses the ``static approximation'' to the nucleon form factor, Eq.\,\eqref{staticexchange}, as with the contact-interaction herein, then $P_1^{p,m}\equiv 0$.

It is plain from Eq.\,\eqref{dvuvF1result} that $d_v/u_v=0$ in the absence of axial-vector diquark correlations; i.e., in scalar-diquark-only models of the nucleon.  Furthermore, Eq.\,\eqref{dvuvF1result} produces $d_v/u_v=0.05$, $F_2^n/F_2^p=0.30$, using the case-II solution in Ref.\,\cite{Mineo:2002bg}, which is fully consistent with Fig.\,5 therein.

\begin{figure}[t]
\begin{centering}
\includegraphics[clip,width=0.45\textwidth]{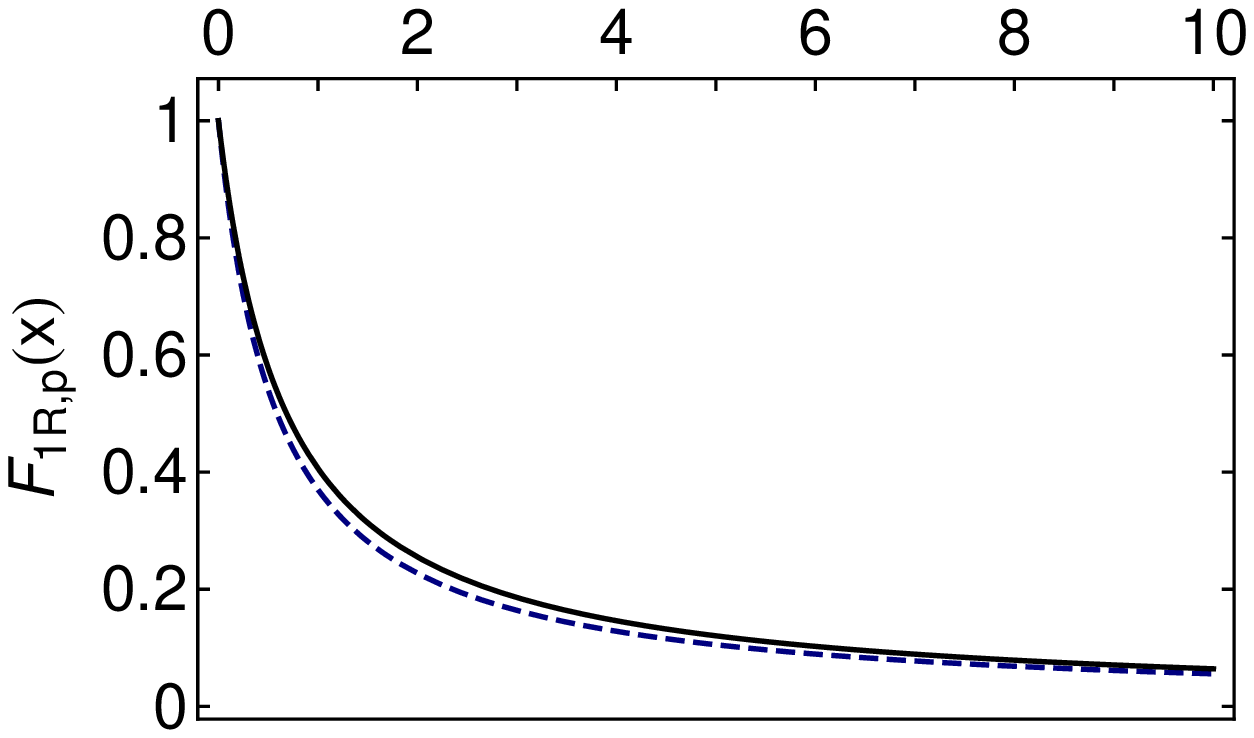}
\vspace*{-4.5ex}

\includegraphics[clip,width=0.45\textwidth]{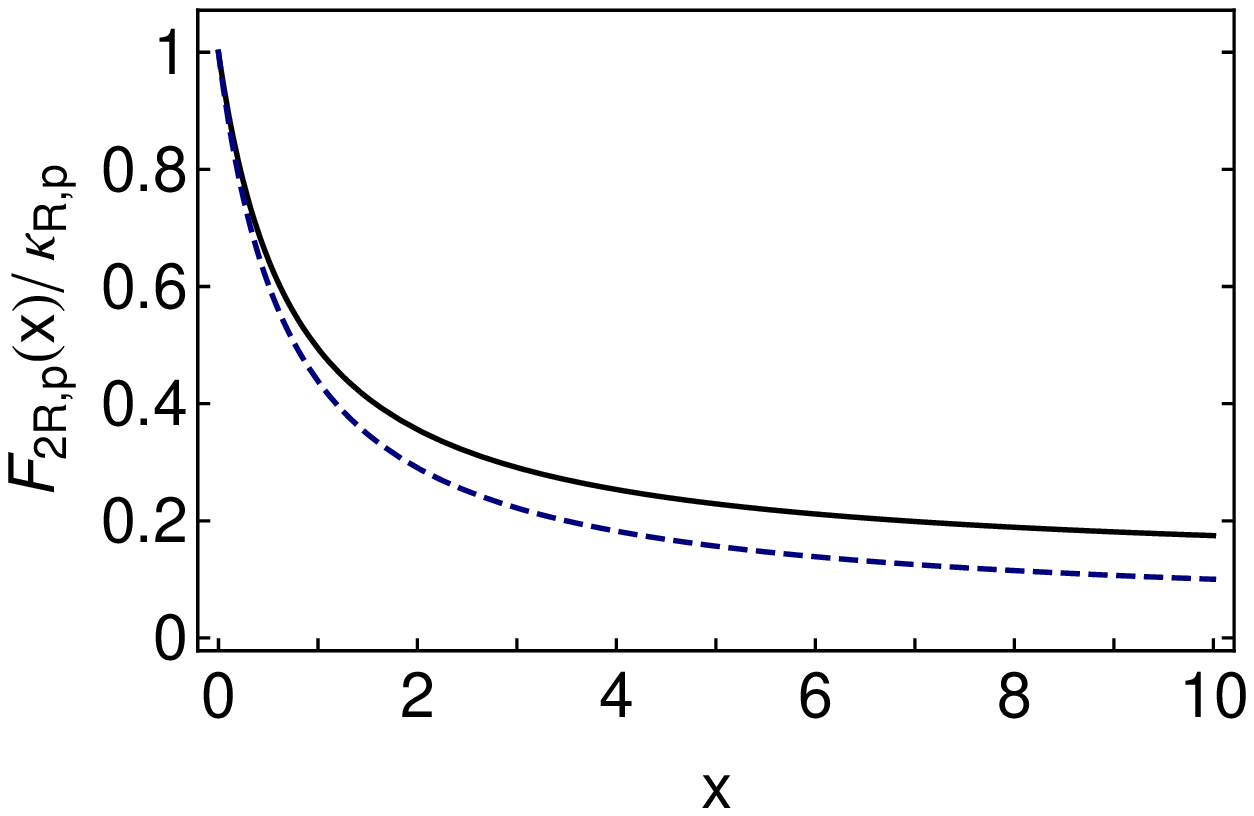}
\end{centering}
\caption{\label{fig:RoperplusF1F2} Comparison of charged-Roper and proton Dirac (upper panel) and Pauli (lower panel) form factors, as a function of $x=Q^2/m_N^2$: \emph{Solid curve} -- Roper; and \emph{dashed-curve} -- proton.  All results obtained using the contact-interaction, and hence a dressed-quark mass-function and diquark Bethe-Salpeter amplitudes that are momentum-independent.}
\end{figure}

\begin{table}[t]
\caption{Probabilities described after Eq.\,\eqref{dvuvF1result}, from which one may compute the evolution-invariant $x=1$ value of the structure function ratio.
\label{Table:DISx1}
}
\begin{center}
\begin{tabular*}
{\hsize}
{
l@{\extracolsep{0ptplus1fil}}
|c@{\extracolsep{0ptplus1fil}}
c@{\extracolsep{0ptplus1fil}}
c@{\extracolsep{0ptplus1fil}}
c@{\extracolsep{0ptplus1fil}}
c@{\extracolsep{0ptplus1fil}}}\hline
 & $P_1^{p,s}$ & $P_1^{p,a}$ & $P_1^{p,m}$ & $\frac{d_v}{u_v}$ & $\frac{F_2^n}{F_2^p}$ \\\hline
\mbox{M=constant} &  0.78 & 0.22 & 0\rule{1.2em}{0ex} & 0.18 & 0.41\\
$M(p^2)$ & 0.60 & 0.25 & 0.15 & 0.28 & 0.49\\\hline
\end{tabular*}
\end{center}
\end{table}

Using the probabilities derived from Table~\ref{Table:FE}B, one obtains the first row in Table~\ref{Table:DISx1},
%
%
whilst the second row is drawn from Ref.\,\protect\cite{Cloet:2008re}.  (Here we correct an error in Ref.\,\cite{Holt:2010vj}, which inadvertently interchanged $2\leftrightarrow 1$ in evaluating the $P_1^{p,a}$ contribution.) Both rows in Table~\ref{Table:DISx1} are consistent with $d_v/u_v= 0.23\pm 0.09$ (90\% confidence level, $F_2^n/F_2^p = 0.45 \pm 0.08$) inferred recently via consideration of electron-nucleus scattering at $x>1$ \cite{Hen:2011rt}.  On the other hand, this is also true of the result obtained through a naive consideration of the isospin and helicity structure of a proton's light-front quark wave function at $x\sim 1$, which leads one to expect that $d$-quarks are five-times less likely than $u$-quarks to possess the same helicity as the
proton they comprise; viz., $d_v/u_v=0.2$ \cite{Farrar:1975yb}.  Plainly, contemporary experiment-based analyses do not provide a particularly discriminating constraint.  Future experiments with a tritium target could help \cite{Holt:2010zz}.

\section{Nucleon$\,\to\,$Roper Transition and Roper Elastic}
\label{sec:transitionelastic}
A computation of the nucleon-to-Roper transition form factors must be performed in conjunction with that of the Roper elastic form factors.  They are connected via orthonormalisation: the Roper is orthogonal to the nucleon, which means $F_{1*}(Q^2=0)=0$ for both the charged and neutral channels; and the canonical normalisation of the Roper Faddeev amplitude is fixed by setting $F_{1R^+}(Q^2=0)=1$.  The transition is calculated with the kinematic arrangements:
\begin{equation}
\label{transitionkinematics}
P_f^2 = -m_R^2\,, \; P_i^2= -m_N^2\,, \; m_R^2 - m_N^2 + 2 P_i\cdot Q + Q^2 =0\,,
\end{equation}
from the transition current expressed by the diagrams in Fig.\,\ref{fig:current}, which are as explained in App.\,\ref{App:current} except that the final baryon, $\Psi_f$, is the Roper resonance.  These considerations lead to the modifications described in App.\,\ref{App:transitioncurrent}.

Note that in connection with all form factors involving the Roper resonance, we only report results obtained with our symmetry-preserving treatment of the contact interaction.  This is a first step.  Based on the information in Sec.\,\ref{sec:nucleonelastic}, we anticipate that a momentum-dependent interaction will produce Roper-related form factors that are similar for $Q^2 \lesssim 0.5\,{\rm GeV}^2$ but softer at larger momentum scales.

\subsection{Roper Faddeev amplitude}
The Faddeev amplitude for the Roper resonance in Table~\ref{Table:FE}B, whose origin is explained in Apps.\,\ref{sec:Faddeev}, \ref{App:transitioncurrent}, contrasts strikingly with that of the nucleon and suggests a fascinating new possibility for the structure of the Roper's dressed-quark core.  To explain this remark, we focus first on the nucleon, whose Faddeev amplitude describes a ground-state that is dominated by its scalar diquark component (78\%).  The axial-vector component is significantly smaller but nevertheless important.  This heavy weighting of the scalar diquark component persists in solutions obtained with more sophisticated Faddeev equation kernels (see, e.g., Table~2 in Ref.\,\cite{Cloet:2008re}).  From a perspective provided by the nucleon's parity partner and the radial excitation of that state, in which the scalar and axial-vector diquark probabilities are \cite{Roberts:2011ym} 51\%-49\% and 43\%-57\%, respectively, the scalar diquark component of the ground-state nucleon actually appears to be unnaturally large.

\begin{table}[t]
\caption{Row~1: Roper results computed herein with the contact interaction, whose input is presented in Table~\protect\ref{Table:FE}.
Row~2: Related contact-interaction nucleon results repeated for ease of comparison.
%
Rows~3, 4: Analogous results obtained with a model dressed-quark anomalous magnetic moment, Sec.\,\ref{sec:HAMM}.
\label{tab:Roperstatic}
}
\begin{center}
\begin{tabular*}
{\hsize}
{
l@{\extracolsep{0ptplus1fil}}
c@{\extracolsep{0ptplus1fil}}
c@{\extracolsep{0ptplus1fil}}
c@{\extracolsep{0ptplus1fil}}
c@{\extracolsep{0ptplus1fil}}
c@{\extracolsep{0ptplus1fil}}
c@{\extracolsep{0ptplus1fil}}}
    & $r_{1}^{R+} M_N$ & $r_{2}^{R^+} M_N$ & $r_{1}^{R^0} M_N$ & $r_{2}^{R^0} M_N$ & $\kappa_{R^+}$& $\kappa_{R^0}$\\\hline
Roper & 2.96 & 2.66 & 0.81 & 3.19 & 0.61 & -0.61 \\
Nucleon & 3.19 & 2.84 & 1.21 & 3.19 & 1.02 & -0.92\\\hline
Roper$_{\rm QAMM}$  & 3.29 & 3.90 & 0.22 & 3.46  & 1.75 & -1.20\\
Nucleon$_{\rm QAMM}$ & 3.41 & 4.00 & 0.55 & 3.85 & 1.68 & -1.24 \\\hline
\end{tabular*}
\end{center}
\end{table}

One can nevertheless understand the structure of the nucleon.  As with so much else, the composition of the nucleon is intimately connected with dynamical chiral symmetry breaking.  In a two-color version of QCD, the scalar diquark is a Goldstone mode, just like the pion \cite{Roberts:1996jx}.  (This is a long-known result of Pauli-G\"ursey symmetry.)  A memory of this persists in the three-color theory and is evident in many ways.  Amongst them, through a large value of the canonically normalized Bethe-Salpeter amplitude and hence a strong quark$+$quark$-$diquark coupling within the nucleon.  (A qualitatively identical effect explains the large value of the $\pi N$ coupling constant.) There is no such enhancement mechanism associated with the axial-vector diquark.  Therefore the scalar diquark dominates the nucleon.

\begin{figure}[t]
\begin{centering}
\includegraphics[clip,width=0.45\textwidth]{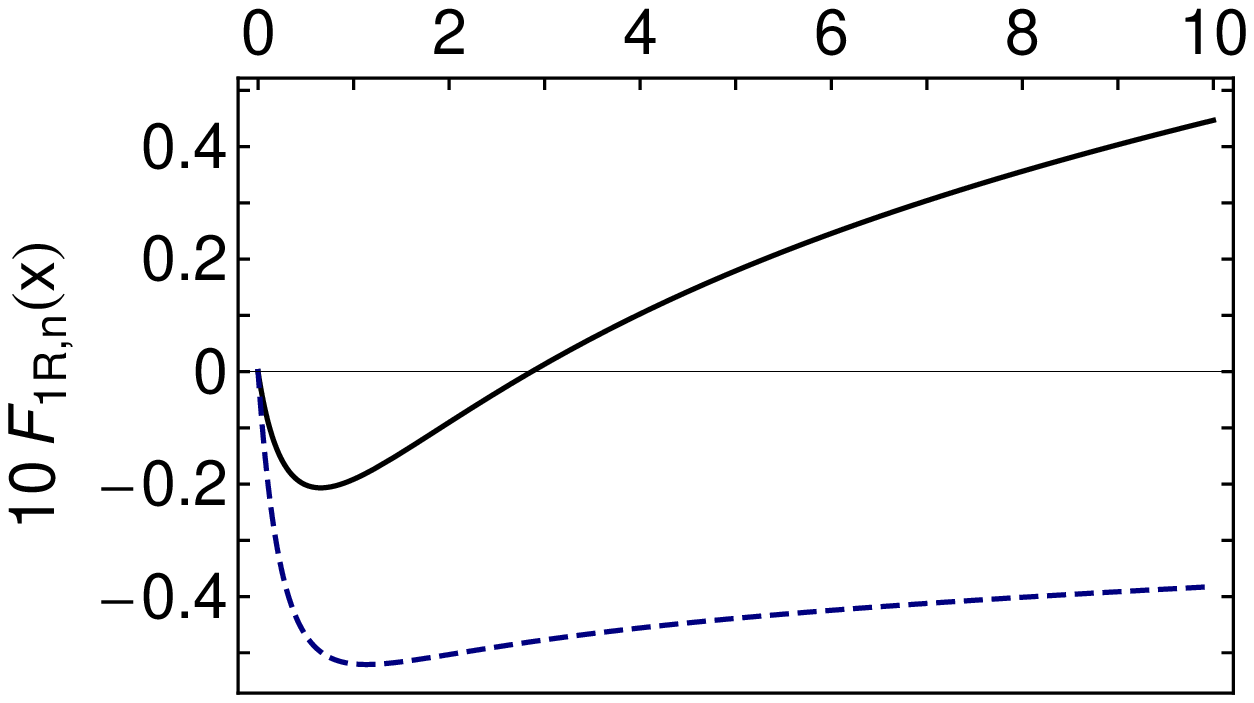}
\vspace*{-4.5ex}

\includegraphics[clip,width=0.45\textwidth]{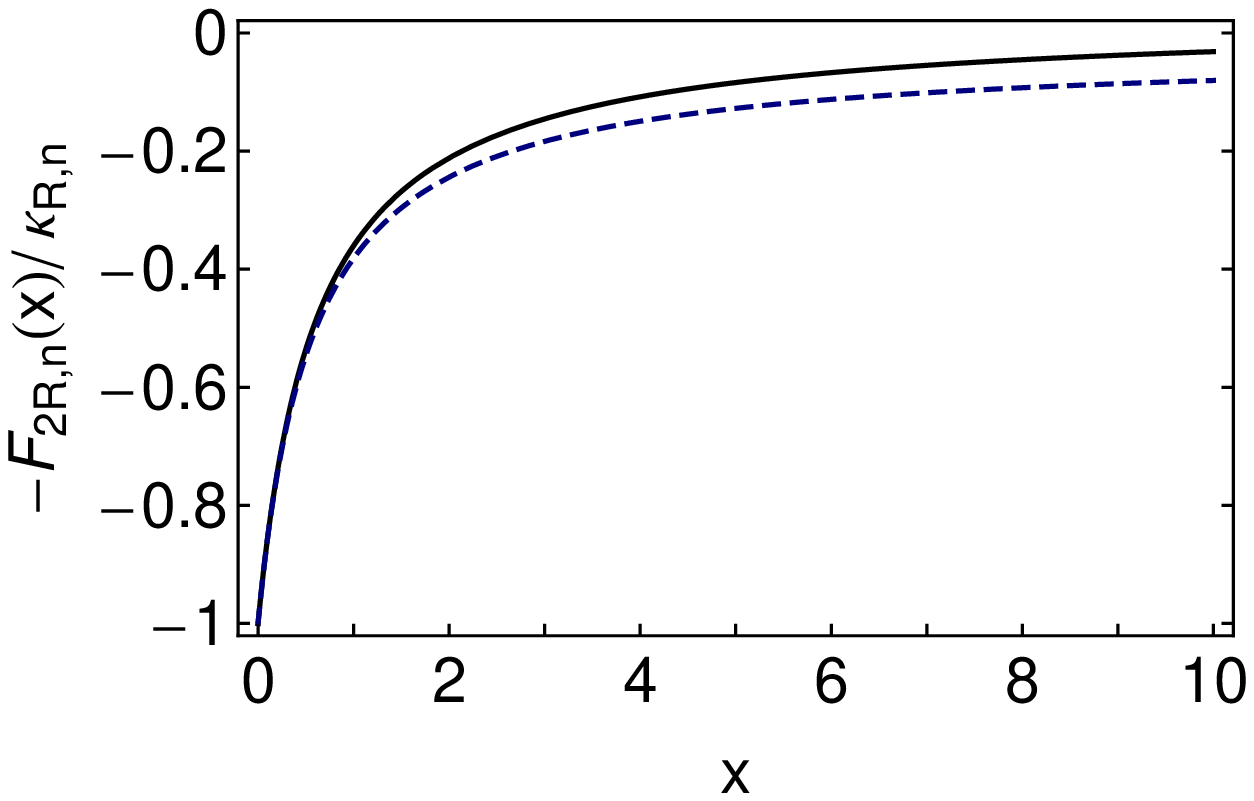}
\end{centering}
\caption{\label{fig:RoperzeroF1F2} Comparison of neutral-Roper and neutron Dirac (upper panel) and Pauli (lower panel) form factors, as a function of $x=Q^2/m_N^2$: \emph{Solid curve} -- neutral-Roper; and \emph{dashed-curve} -- neutron.  All results obtained using the contact-interaction, and hence a dressed-quark mass-function and diquark Bethe-Salpeter amplitudes that are momentum-independent.}
\end{figure}

With the Faddeev equation treatment described herein, the effect on the Roper is dramatic: orthogonality of the ground- and excited-states forces the Roper to be constituted almost entirely (81\%) from the axial-vector diquark correlation.  It is important to check whether this outcome survives with a Faddeev equation kernel built from a momentum-dependent interaction.

\begin{figure}[t]
\begin{centering}
\includegraphics[clip,width=0.455\textwidth]{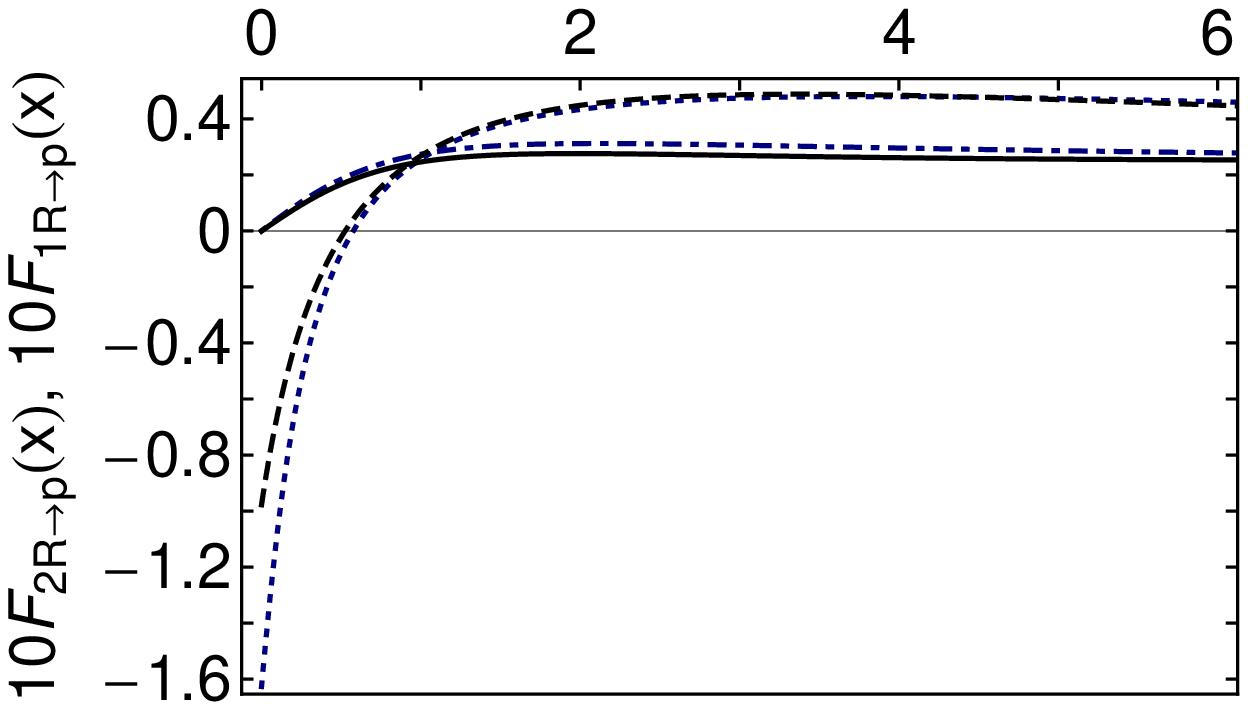}\hspace*{0.2em}
\vspace*{-4.5ex}

\includegraphics[clip,width=0.45\textwidth]{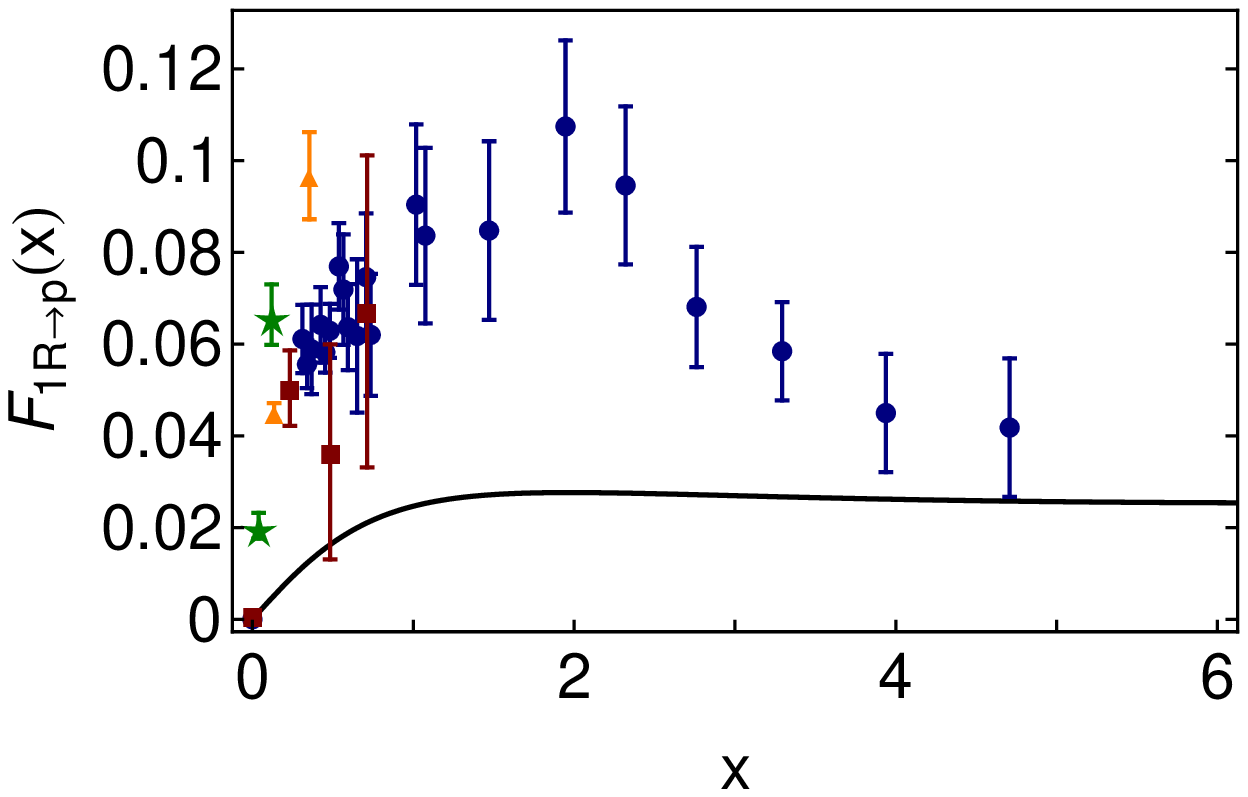}
\end{centering}
\caption{\label{fig:NRoper1}
\emph{Upper panel} -- $F_{1\ast}$ (solid and dot-dashed with dressed-quark anomalous magnetic moment, Sec.\,\protect\ref{sec:HAMM}) and $F_{2\ast}$ (dashed and dotted with dressed-quark anomalous magnetic moment) as a function of $x=Q^2/m_N^2$, computed using the framework described herein.
\emph{Lower panel} -- Computed form of $F_{1\ast}(x)$ compared with available data \protect\cite{Dugger:2009pn,Aznauryan:2009mx,Aznauryan:2011td}.
The squares, triangles and stars are preliminary results \protect\cite{Lin:2011da} from a simulation of $N_f=2+1$ lattice-QCD at, respectively, $m_\pi^2/m_{\pi{\rm expt.}}^2 \simeq 8\,$, $10$, $40$.}
\end{figure}

\subsection{Roper elastic}
The Roper mass and Faddeev amplitude in Table~\ref{Table:FE}B produce the radii and anomalous magnetic moments in Table~\ref{tab:Roperstatic} and the elastic form factors depicted in Figs.\,\ref{fig:RoperplusF1F2}, \ref{fig:RoperzeroF1F2}.  Notwithstanding the markedly different internal structure, the Roper elastic form factors are similar to those of the nucleon, both in magnitude and $Q^2$-evolution.

\begin{figure}[t]
\begin{centering}
\includegraphics[clip,width=0.45\textwidth]{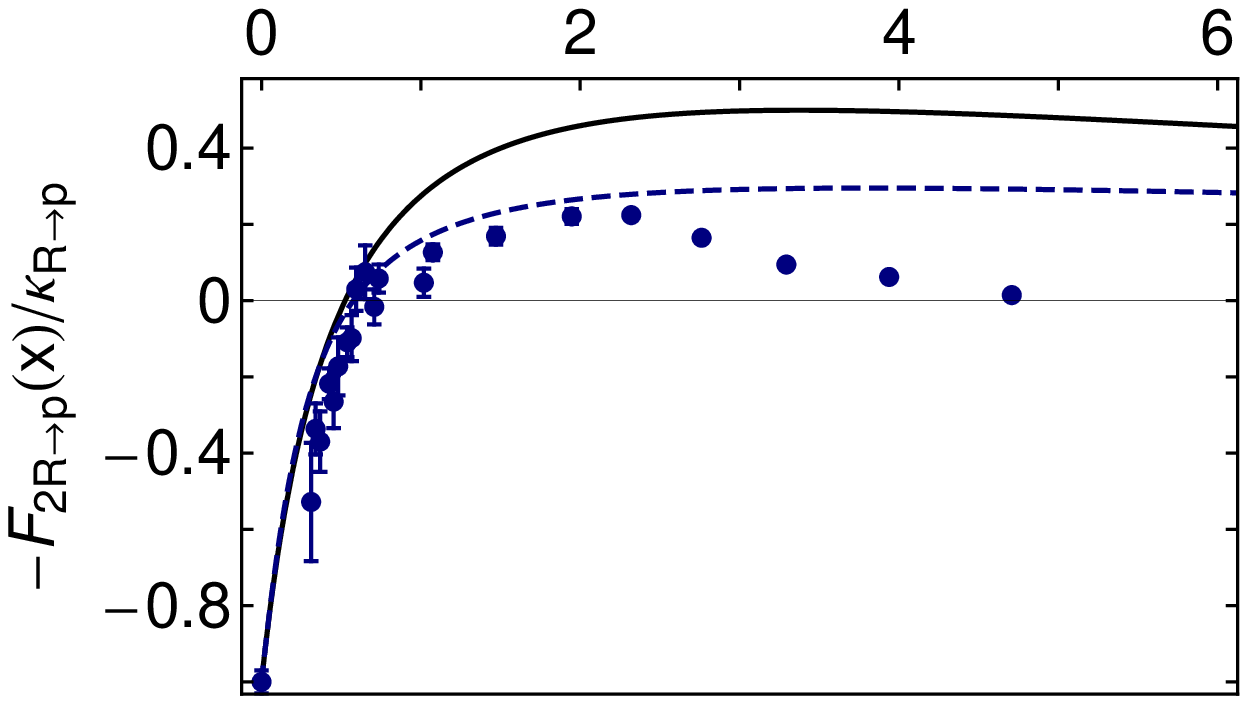}
\vspace*{-8.0ex}

\includegraphics[clip,width=0.45\textwidth]{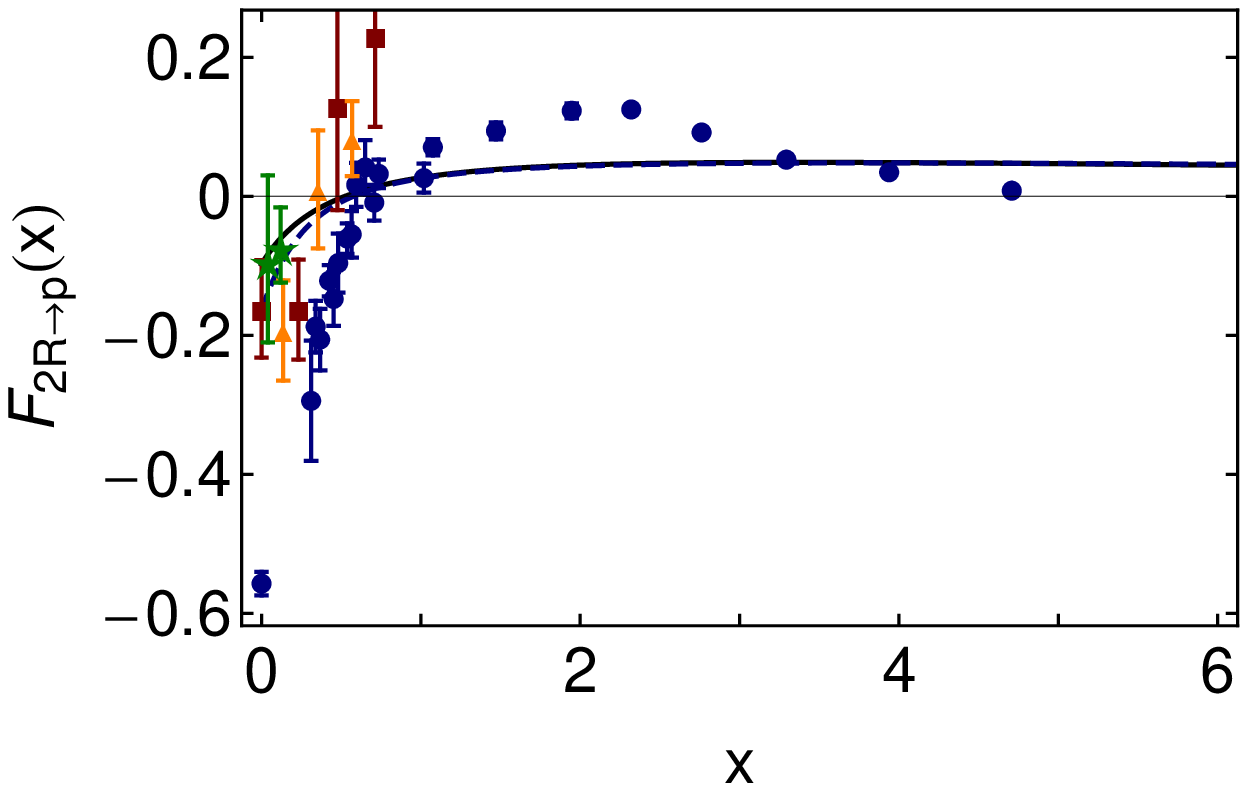}
\end{centering}
\caption{\label{fig:NRoper2}
Comparison between $F_{2\ast}(x)$ computed using the framework described herein and available data \protect\cite{Dugger:2009pn,Aznauryan:2009mx,Aznauryan:2011td}, with $x=Q^2/m_N^2$.
\emph{Upper panel} -- normalised to unity at $x=0$; and \emph{lower panel}, as computed.
In both panels the dashed curve was computed with a model for the dressed-quark anomalous electromagnetic moment, Sec.\,\protect\ref{sec:HAMM}.
The squares, triangles and stars are preliminary results from a simulation of $N_f=2+1$ lattice-QCD at, respectively, $m_\pi^2/m_{\pi{\rm expt.}}^2 \simeq 8\,$, $10$, $40$ \protect\cite{Lin:2011da}.
}
\end{figure}

The exception is the Dirac form factor of the neutral Roper, which exhibits a zero at $Q^2 \simeq 3 m_N^2$.  This behaviour derives from a constructive interference between Diagrams~2 and 3 in Fig.\,\ref{fig:current} that, with increasing $Q^2$, sums to overwhelm the always-negative contribution from Diagram~1.  As $Q^2$ increases, the dominant contributions expressed by Diagrams~2 and 3 are associated with a photon scattering from the positively-charged $[ud]$ and $\{ud\}$ correlations, whereas Diagram~1 is alone in measuring only a negative charge; i.e., that of the $d$-quark.  Ultimately, therefore, suppression of the scalar-diquark component in the Roper is responsible for the zero in $F_{1R^0}$ at $Q^2>0$.

\subsection{Transition}
In Figs.\,\ref{fig:NRoper1}, \ref{fig:NRoper2} we depict the charged-Roper$\,\to\,$proton transition form factors computed using our treatment of the contact interaction.
The calculated form factors underestimate the data on the domain $0<Q^2<3\,$GeV$^2$ and are very probably too hard.  Both of these defects are natural given that we have: deliberately omitted effects associated with a meson cloud in the Faddeev kernel and the current; and used a contact interaction.

On the other hand, the results are qualitatively in agreement with the trend apparent in available data and reproduce the zero in $F_{2\ast}(Q^2)$ at $Q^2 \simeq 0.5\,m_N^2$ without fine tuning.  These are meaningful successes given that they are features derived only from that which we consider to be the Roper's dressed-quark core.

As shown in the figures, lattice-QCD results are also available for these form factors  \protect\cite{Lin:2011da}.  They have roughly the same magnitude as the experimental data.  In contrast to earlier simulations of quenched-QCD, these $N_f=2+1$ results also support the presence of a zero in $F_{2\ast}$.

\begin{figure}[t]
\begin{centering}
\includegraphics[clip,width=0.45\textwidth]{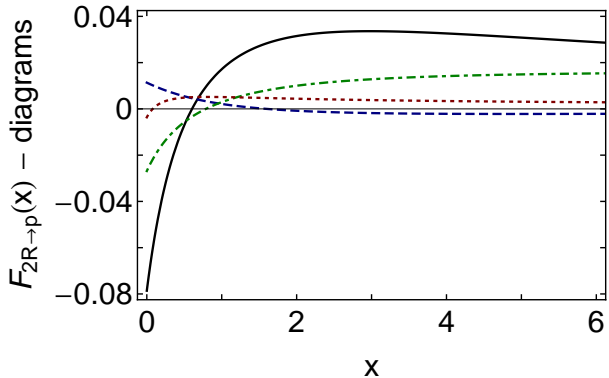}
\end{centering}
\caption{\label{fig:NRoperDiagrams}
Separation of $F_{2\ast}(x)$ into contributions from different diagrams, with $x=Q^2/m_N^2$:
\emph{solid} -- photon on $u$-quark with scalar diquark spectator;
\emph{dashed} -- photon on scalar diquark with $u$-quark spectator;
\emph{dot-dashed} -- photon on axial-vector diquark with quark spectator;
\emph{dotted} -- photon-induced transition between scalar and axial-vector diquarks with $u$-quark spectator.  N.B.\ Owing to Eq.\,\protect\eqref{isospinFF}, there is no contribution involving an axial-vector diquark spectator.}
\end{figure}

In Fig.\,\ref{fig:NRoperDiagrams} we display the separate contributions from each diagram represented by the current in Fig.\,\ref{fig:current}.  Whilst Diagram~1 with a scalar diquark bystander is plainly dominant, a significant contribution is also received from Diagram~2 with a photon probing the structure of the axial-vector diquark correlations.  The form factor is negative at $Q^2=0$ owing to orthogonality, which produces $s_R s_N < 0$, and passes through zero because of the zero in the Roper's Faddeev amplitude, which is characteristic of a radial excitation.

\begin{figure}[t]
\begin{centering}
\includegraphics[clip,width=0.45\textwidth]{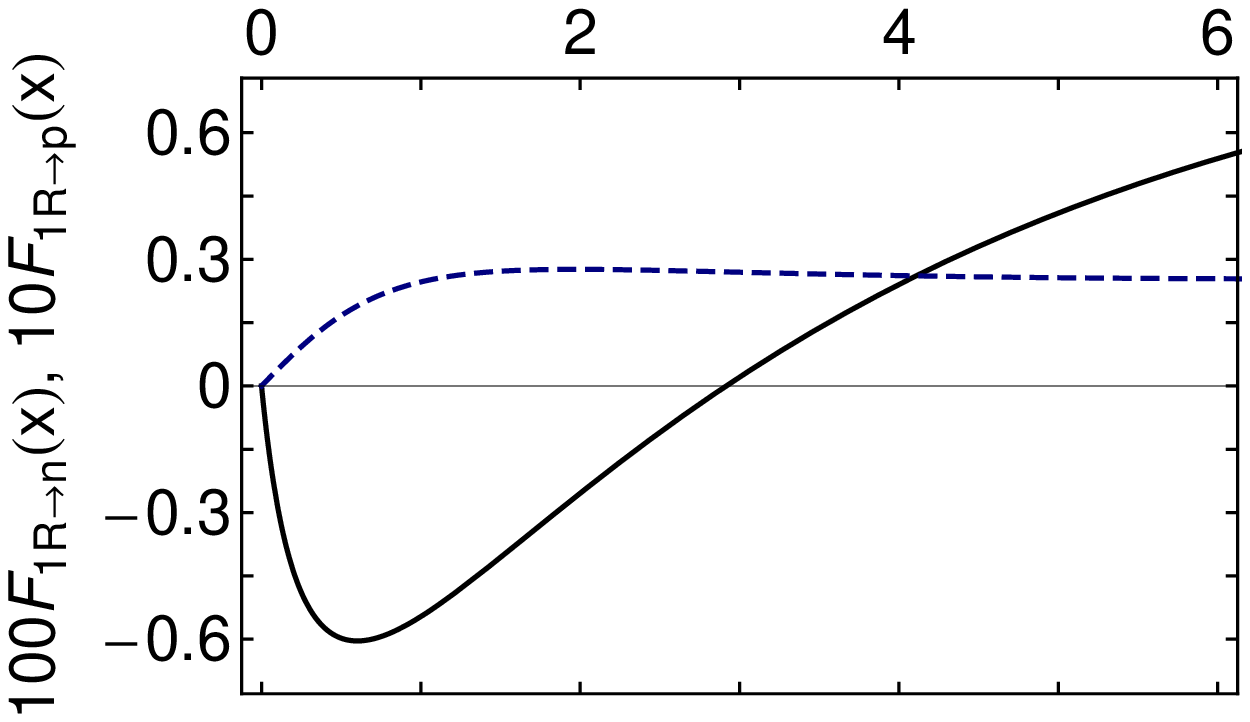}
\vspace*{-9.5ex}

\includegraphics[clip,width=0.45\textwidth]{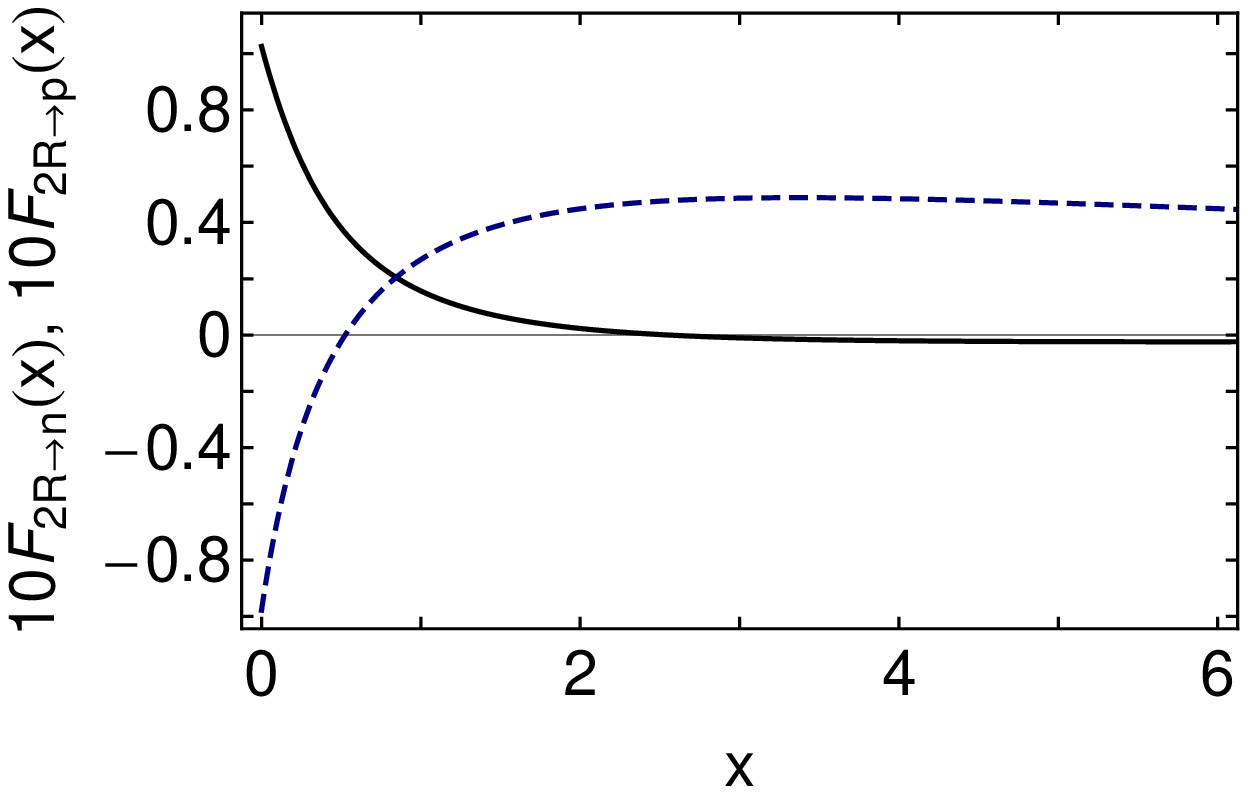}
\end{centering}
\caption{\label{fig:NRoperNeutron}
\emph{Upper panel} -- $F_{1R^0\to n}$ (solid) as a function of $x=Q^2/m_N^2$ compared with $F_{1R^+\to p}$ (dashed), computed using the framework described herein.
\emph{Lower panel} -- Analogue for $F_{2 R^0\to n}$.}
\end{figure}

Figure~\ref{fig:NRoperNeutron} depicts the neutral-Roper$\,\to\,$neutron transition form factors. Each possesses a zero at $Q^2\simeq 3 m_N^2$; the Dirac form factor is an order-of-magnitude smaller than its analogue in the charged-Roper transition; and regarding $F_{2R^0\to n}$ cf.\ $F_{2R^+\to p}$, in the neighbourhood of $Q^2=0$ the similar magnitude but opposite sign is consistent with available data \cite{Nakamura:2010zzi}.

\section{Anomalous magnetic moments}
\label{sec:HAMM}
It is noticeable from the lower panel of Fig.\,\ref{fig:NRoper2} that the magnitude of $F_{2\ast}(Q^2=0)$ is underestimated in our framework: $-0.1$ cf.\ experiment \cite{Dugger:2009pn}, $-0.56 \pm  0.02$.  A similar but smaller deficit is apparent in our computed nucleon anomalous electromagnetic moments, Table~\ref{tab:nucleonstatic}.  In this connection it is interesting to explore the effect produced by the dressed-quark anomalous electromagnetic moment, which is produced by DCSB \cite{Chang:2010hb} and is known to have a material impact on the nucleons' Pauli form factors \cite{Chang:2011tx}.

To this end we modified the quark-photon coupling as described in App.\,\ref{App:AMM} and recomputed all the form factors described above.  Some results for the nucleon are summarised in the last row of Table~\ref{tab:nucleonstatic}: in each case, inclusion of the dressed-quark anomalous magnetic moment produces a significant improvement in the comparison with data.  A similar comparison is made for the Roper in Table~\ref{tab:Roperstatic}.

Results for the Roper$\,\to\,$proton transition form factor are included in Figs.\,\ref{fig:NRoper1}, \ref{fig:NRoper2}.  Inclusion of a dressed-quark anomalous electromagnetic moment has a pronounced effect on $F_{2\ast}$, which moves the result a little closer to experiment: $F_{2\ast}(Q^2=0) = -0.1\to -0.16$ cf.\ experiment \cite{Dugger:2009pn} $-0.56 \pm  0.02$.  It does not, however, compensate sufficiently for the absence of meson-cloud effects.

\begin{figure}[t]
\begin{centering}
\includegraphics[clip,width=0.462\textwidth]{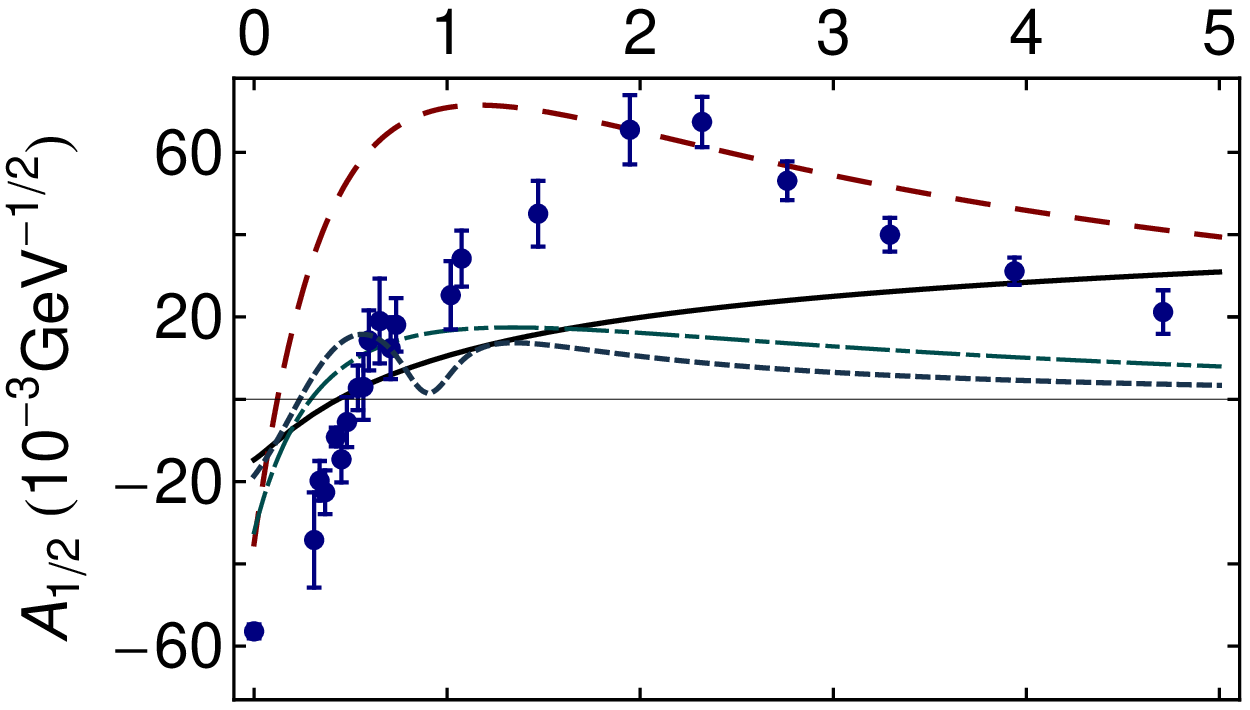}\hspace*{0.6em}
\vspace*{-9.4ex}

\includegraphics[clip,width=0.45\textwidth]{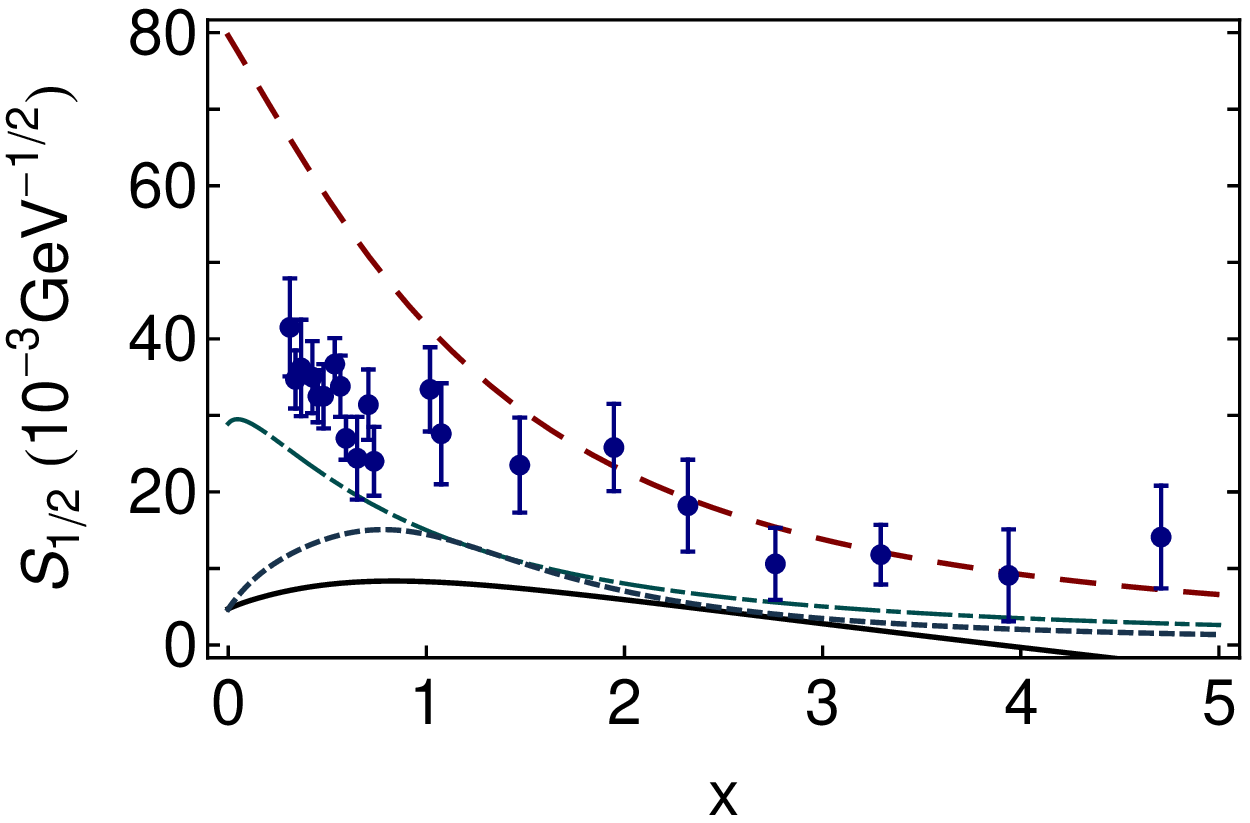}
\end{centering}
\caption{\label{fig:RoperHelicity}
Helicity amplitudes for the $\gamma^\ast p \to P_{11}(1440)$ transition, with $x=Q^2/m_N^2$:
$A_{1/2}$ (upper panel); and $S_{1/2}$ (lower panel).
Solid curves -- computed using the treatment of the contact interaction described herein, including the dressed-quark anomalous magnetic moment (App.\,\protect\ref{App:AMM});
dashed curves -- the light-front constituent quark model results from Ref.\,\protect\cite{Aznauryan:2007ja};
long-dash-dot curves -- the light-front constituent quark model results from Ref.\,\protect\cite{Cardarelli:1996vn};
short-dashed curves -- our smooth fit to the bare form factors inferred in Ref.\,\protect\cite{Suzuki:2010yn,JuliaDiaz:2009ww,LeePrivate:2011};
and data -- Refs.\,\protect\cite{Dugger:2009pn,Aznauryan:2009mx,Aznauryan:2011td}.
}
\end{figure}

\section{Meson Cloud}
\label{sec:mesoncloud}
In Fig.\,\ref{fig:RoperHelicity} we draw the helicity amplitudes for the $\gamma^\ast p \to P_{11}(1440)$ transition. 
They may be computed from the transition form factors in Eq.\,\eqref{NRtransition}:
\begin{subequations}
\begin{eqnarray}
A_{\frac{1}{2}}(Q^2) & =& c(Q^2) \left[ F_{1\ast}(Q^2)+ F_{2\ast}(Q^2) \right],\\
\nonumber S_{\frac{1}{2}}(Q^2) & = & -\frac{q_{\rm CMS}}{\surd 2} c(Q^2)
\left[ - F_{1\ast}(Q^2) \frac{m_R+m_N}{Q^2} \right. \\
&& \left. + \frac{F_{2\ast}(Q^2)}{m_R+m_N}  \right],
\end{eqnarray}
\end{subequations}
with
\begin{equation}
c(Q^2) = \left[ \frac{ \pi \alpha Q^2_-}{m_R m_N K} \right]^{\frac{1}{2}},\;
q_{\rm CMS} = \frac{\sqrt{Q_-^2 Q_+^2}}{2 m_R},
\end{equation}
where $Q_{\pm}^2 = Q^2 + (m_R \pm m_N)^2$, $K= (m_R^2-m_N^2)/(2 m_R)$, and $\alpha$ is QED's fine structure constant.

In addition to our own computation, Fig.\,\ref{fig:RoperHelicity} displays results obtained using a light-front constituent-quark model \cite{Aznauryan:2007ja}, which employed a constituent-quark mass of $0.22\,$GeV and identical momentum-space harmonic oscillator wave functions for both the nucleon and Roper (width$\,=0.38\,$GeV) but with a zero introduced for the Roper, whose location was fixed by an orthogonality condition.
The quark mass is smaller than the DCSB-induced value we determined from the gap equation (see Table~\ref{Table:FE}) but a more significant difference is the choice of spin-flavour wave functions for the nucleon and Roper.  In Ref.\,\cite{Aznauryan:2007ja} they are simple $SU(6)\times O(3)$ $S$-wave states in the three-quark centre-of-mass system, in contrast to the markedly different spin-flavour structure produced by our Faddeev equation analysis of these states, Table~\ref{Table:FE}B.

Owing to this, in Fig.\,\ref{fig:RoperHelicity} we also display the light-front quark model results from Ref.\,\cite{Cardarelli:1996vn}.  It is stated therein that large effects accrue from ``configuration mixing;'' i.e., the inclusion of $SU(6)$-breaking terms and high-momentum components in the wave functions of the nucleon and Roper.  In particular, that configuration mixing yields a marked suppression of the calculated helicity amplitudes in comparison with both relativistic and non-relativistic results based on a simple harmonic oscillator \emph{Ansatz} for the baryon wave functions, as used in Ref.\,\cite{Aznauryan:2007ja}.

There is also another difference; namely, Ref.\,\cite{Cardarelli:1996vn} employs Dirac and Pauli form factors to describe the interaction between a photon and a constituent-quark \cite{Cardarelli:1995dc}.
As apparent in Fig.\,2 of Ref.\,\cite{Cardarelli:1996vn}, they also have a noticeable impact, providing roughly half the suppression on $0.5\lesssim Q^2/{\rm GeV}^2 \lesssim 1.5$.  The same figure also highlights the impact on the form factors of high-momentum tails in the nucleon and Roper wave functions.

In reflecting upon constituent-quark form factors, we note that the interaction between a photon and a dressed-quark in QCD is not simply that of a Dirac fermion \cite{Ball:1980ay,Curtis:1990zs,Alkofer:1993gu,Frank:1994mf,Roberts:1994hh,Maris:1999bh,Chang:2010hb}.  Moreover, the interaction of our dressed-quark with the photon is also modulated by form factors, see Apps.\,\ref{sec:WTI}, \ref{App:AMM}.  On the other hand, the purely phenomenological form factors in Refs.\,\cite{Cardarelli:1995dc,Cardarelli:1996vn} are inconsistent with a number of constraints that apply to the dressed-quark-photon vertex in quantum field theory; e.g., the dressed-quark's Dirac form factor should approach unity with increasing $Q^2$ and neither its Dirac nor Pauli form factors may possess a zero.  Notwithstanding these observations, the results from Ref.\,\cite{Cardarelli:1996vn} are more similar to ours than those in Ref.\,\cite{Aznauryan:2007ja}.

Helicity amplitudes can also be computed using EBAC's dynamical coupled-channels framework \cite{Matsuyama:2006rp}.  In this approach, one imagines that a Hamiltonian is defined in terms of bare baryon states and bare meson-baryon couplings; the physical amplitudes are computed by solving coupled-channels equations derived therefrom; and the parameters characterising the bare states are determined by requiring a good fit to data.
In connection with the $\gamma^\ast p \to P_{11}(1440)$ transition, results are available for both helicity amplitudes \cite{Suzuki:2010yn,JuliaDiaz:2009ww,LeePrivate:2011}.  The associated bare form factors are reproduced in Fig.\,\ref{fig:RoperHelicity}: for $Q^2<1.5\,$GeV$^2$ we depict a smooth interpolation; and for larger $Q^2$ an extrapolation based on perturbative QCD power laws ($A_{\frac{1}{2}}\sim 1/Q^3 \sim S_{\frac{1}{2}}$).


The bare form factors are evidently similar to the results obtained herein and in Ref.\,\cite{Cardarelli:1996vn}: both in magnitude and $Q^2$-evolution.  Regarding the transverse amplitude, Ref.\,\cite{Suzuki:2010yn} argues that the bare component plays an important role in changing the sign of the real part of the complete amplitude in the vicinity of $Q^2=0$.  In this case the similarity between the bare form factor and the results obtained herein is perhaps most remarkable -- e.g., the appearance of the zero in $A_{\frac{1}{2}}$, and the $Q^2=0$ magnitude of the amplitude (in units of $10^{-3}\,{\rm GeV}^{-1/2}$)
\begin{equation}
\begin{array}{lcccc}
                    & \mbox{Ref.\,\protect\cite{Aznauryan:2007ja}} &
                    \mbox{Ref.\,\protect\cite{Cardarelli:1996vn}} &
                    \mbox{Ref.\,\protect\cite{Suzuki:2010yn,JuliaDiaz:2009ww,LeePrivate:2011}} &
                                        \mbox{contact} \\
A_{\frac{1}{2}}(0)   & -35.1 & -32.3 & -18.6 & -16.3
\end{array}.
\end{equation}
These similarities strengthen support for an interpretation of the bare-masses, -couplings, etc.,  inferred via coupled-channels analyses, as those quantities comparable with hadron structure calculations that exclude the meson-baryon coupled-channel effects which are determined by multichannel unitarity conditions.

An additional remark is valuable in this connection.  EBAC computes electroproduction form factors at the resonance pole in the complex plane and hence they are complex-valued functions.  Whilst this is consistent with the standard theory of scattering \cite{JRTaylor:1972}, it differs markedly from phenomenological approaches that use a Breit-Wigner parametrisation of resonant amplitudes in fitting data.  As concerns the $\gamma^\ast p \to P_{11}(1440)$ transition, the real parts of EBAC's complete amplitudes are qualitatively similar to the results in Refs.\,\cite{Aznauryan:2008pe,Dugger:2009pn,Aznauryan:2009mx,Aznauryan:2011td} but EBAC's amplitudes also have sizeable imaginary parts.  This complicates a direct comparison between theory and extant data.

\section{Epilogue}
\label{sec:summary}
We computed form factors for elastic electromagnetic nucleon and Roper scattering and nucleon$\,\to\,$Roper transitions using a Poincar\'e-covariant, symmetry-preserving DSE-treatment of a vector$\,\times\,$vector contact-interaction.
Within this internally-consistent framework current-conservation is assured and we obtain:
a dressed-quark that is described by a momentum-independent mass-function but whose computed interaction with the photon is described by a $Q^2$-dependent vertex;
scalar and axial-vector diquark correlations (constituted from dressed-quarks) whose Bethe-Salpeter amplitudes are independent of constituent relative momentum but whose interactions with the photon are described by calculated $Q^2$-dependent form factors;
and baryons, whose nontrivial spin-flavour structure is determined from the solution of a Faddeev equation, which produces a bound-state comprised from dressed-quarks and -diquarks, described by a momentum-independent Faddeev amplitude but whose elastic electromagnetic and transition form factors are $Q^2$-dependent.

We found that the electromagnetic interactions of baryons constituted thus from the contact interaction are typically described by hard form factors.  Although this was to be expected, it is nevertheless important to compute and record the behaviour because this hardness contrasts markedly with results obtained from the momentum-dependent Faddeev amplitudes produced by dressed-quark propagators and diquark Bethe-Salpeter amplitudes with QCD-like momentum-dependence, and with experiment.  Hence the present calculations provide concrete comparisons which support a view that experiment is a sensitive probe of the evolution of the strong interaction's running masses and coupling to infrared momenta, and hence of the long-range behaviour of the $\beta$-function.

In this connection, our analysis of the proton's elastic form factors suggests that the existence, and location if so, of a zero in the ratio of Sachs form factors are strongly influenced by the running of the dressed-quark mass.  Our calculations indicate that a constant mass-function produces a zero at a small value of $Q^2$; a mass-function that is very soft will not produce a zero; and a mass-function which resembles that obtained in the best available DSE- and lattice-QCD studies, produces a zero at a location that is consistent with extant data.  Obtaining a clear experimental answer to the question of whether or not there is a zero, and its location in the latter case, is therefore particularly important.

It is worth reiterating that the diquark correlations, whose properties are computed and employed herein, are composite and fully-interacting.  They must not be confused with the pointlike and sometimes inert degrees-of-freedom used in constituent-quark$+$constituent-diquark potential models of baryons.  Indeed, our analysis showed that the structure and interactions of the diquark correlations play an important role in the development of each baryon form factor.  For example, they are instrumental in producing a zero in the Dirac form factor of the proton's $d$-quark and in determining the ratio of $d$-to-$u$ valence-quark distributions at $x=1$.
It is unsound and misleading to employ a framework in which the correlations are considered as inert and structureless.

We found that the Roper elastic electromagnetic form factors are generally similar to those of the nucleon, both in magnitude and $Q^2$-evolution.  The one exception is the neutral Roper's Dirac form factor, which exhibits a zero at $Q^2\sim 3\,$GeV$^2$.  This outcome owes particularly to the presence of electromagnetically-active diquark correlations.  It is notable in this connection that our treatment of the contact interaction produces a first excitation of the nucleon which is constituted almost entirely (81\%) from axial-vector
diquark correlations.   This is an intriguing possibility that should be checked using a Faddeev equation kernel built from an interaction with QCD-like momentum dependence.

A primary motivation for this study was a desire to correlate nucleon elastic and transition form factors, so that the latter could be considered well-constrained, and then probe further for a connection between the properties of a baryon's dressed-quark core and the bare quantities which feature in modern coupled-channels analyses of resonance electroproduction.  We focussed primarily on the $\gamma^\ast p \to P_{11}(1440)$ transition and obtained form factors that underestimate extant data on the domain $0 < Q^2 < 3\,$GeV$^2$.  This is consistent with having deliberately omitted effects associated with a meson cloud in the Faddeev kernel and
the current.  On the other hand, the results are qualitatively in agreement with the trend of available data; for instance, $F_{2\ast}(Q^2)$ obtained from the dressed-quark core exhibits a zero at $Q^2 \approx 0.5\,m_N^2$.

In Faddeev equation treatments of a baryon's dressed-quark core it is common to find that anomalous electromagnetic moments are underestimated.  This is apparent herein, in connection, too, with transition form factors.  We therefore explored the effect produced by a dressed-quark anomalous electromagnetic moment, whose existence is an essential consequence of DCSB.  We found that with a realistic value for this dressed-quark moment, the magnitudes of hadron magnetic moments are typically increased by $\sim 90$\% and magnetic radii by $\sim 30$\%, and thereafter agree much better with experiment.

As mentioned above, on the domain $0 < Q^2\lesssim 2\,$GeV$^2$ it is widely suspected that the inclusion of effects associated with strong meson-baryon final state interactions -- the so-called meson cloud -- is important in making a realistic comparison between experiment and hadron structure calculations.  We considered this conjecture in the context of the $\gamma^\ast p \to P_{11}(1440)$  helicity amplitudes and found that the bare amplitudes determined via coupled-channels analyses are similar to the form factors produced by our dressed-quark core, both in magnitude and $Q^2$-evolution.  This outcome strengthens support for an interpretation
of the bare-masses, -couplings, etc., inferred via coupled-channels analyses, as those quantities with which the results of hadron structure calculations should directly be compared, if those calculations have knowingly excluded the meson-cloud.

The Roper-related calculations we have described should now be repeated using a momentum dependent interaction that is drawn, as closely as reasonably possible, from the behaviour of QCD.  We expect this to produce form factors that, for $Q^2 \lesssim 0.5\,{\rm GeV}^2$, are similar to those we have obtained from the contact-interaction, but softer at larger momentum scales.  Near term, such computations are achievable within the framework of Ref.\,\cite{Cloet:2008re}, which has provided the basis for many comparisons herein.  Looking further ahead, we anticipate that some priority will be given to the improvement of computational techniques, so that the interaction of Ref.\,\cite{Qin:2011dd}, e.g., can be used directly in the study of transitions to excited states, in analogy with the treatment of ground-state nucleon form factors \cite{Eichmann:2008ef,Eichmann:2011vu,Mader:2011zf}.

\section*{Acknowledgments}
We acknowledge valuable communications with I.~Aznauryan, R.~Gothe, T.-S.\,H.~Lee, H.-W.~Lin, V.~Mokeev, G.~Salm\'e, T.~Sato and S.\,M.~Schmidt.
This work was supported by:
U.\,S.\ Department of Energy, Office of Nuclear Physics, contract no.~DE-AC02-06CH11357;
the University of Adelaide and the Australian Research Council through grant no.~FL0992247;
and Forschungszentrum J\"ulich GmbH.

\appendix
\section{Contact interaction}
\label{sec:contact}
\subsection{Gap equation}
\label{sec:gap}
The starting point for our study is the dressed-quark propagator, which is obtained from the gap equation:
\begin{eqnarray}
\nonumber \lefteqn{S(p)^{-1}= i\gamma\cdot p + m}\\
&&+ \!\! \int \! \frac{d^4q}{(2\pi)^4} g^2 D_{\mu\nu}(p-q) \frac{\lambda^a}{2}\gamma_\mu S(q) \frac{\lambda^a}{2}\Gamma_\nu(q,p) ,\;
\label{gendse}
\end{eqnarray}
wherein $m$ is the Lagrangian current-quark mass, $D_{\mu\nu}$ is the vector-boson propagator and $\Gamma_\nu$ is the quark--vector-boson vertex.  Much is now known about $D_{\mu\nu}$ in QCD \cite{Boucaud:2011ug} and nonperturbative information is accumulating on $\Gamma_\nu$ \cite{Skullerud:2003qu,Chang:2009zb,Chang:2010hb,Chang:2011ei}.  However, this is one of a series of studies undertaken in order to build a stock of material that can be used to identify unambiguous signals in experiment for the pointwise behaviour of: the interaction between light-quarks; the light-quark's mass-function; and other similar quantities.  Whilst these are particular qualities, taken together they can plausibly enable a characterisation of the nonperturbative behaviour of the theory underlying strong interaction phenomena \cite{Aznauryan:2009da,Holt:2010vj,Chang:2011vu}.

We therefore work with the following choice
\begin{equation}
\label{njlgluon}
g^2 D_{\mu \nu}(p-q) = \delta_{\mu \nu} \frac{4 \pi \alpha_{\rm IR}}{m_G^2}\,,
\end{equation}
where $m_G=0.8\,$GeV is a gluon mass-scale typical of the one-loop renormalisation-group-improved interaction introduced in Ref.\,\cite{Qin:2011dd}, and the fitted parameter $\alpha_{\rm IR}/\pi = 0.93$ is commensurate with contemporary estimates of the zero-momentum value of a running-coupling in QCD \cite{Aguilar:2010gm,Boucaud:2010gr}.  Equation~\eqref{njlgluon} is embedded in a rainbow-ladder truncation of the DSEs, which is the leading-order in the most widely used, global-symmetry-preserving truncation scheme \cite{Bender:1996bb}.  This means
\begin{equation}
\label{RLvertex}
\Gamma_{\nu}(p,q) =\gamma_{\nu}
\end{equation}
in Eq.\,(\ref{gendse}) and in the subsequent construction of the Bethe-Salpeter kernels.

One may view the interaction in Eq.\,(\ref{njlgluon}) as being inspired by models of the Nambu--Jona-Lasinio type \cite{Nambu:1961tp}.  However, our treatment is atypical.  It is notable that one typically finds Eqs.\,\eqref{njlgluon}, (\ref{RLvertex}) produce results for low-momentum-transfer observables that are practically indistinguishable from those produced by more sophisticated interactions \cite{GutierrezGuerrero:2010md,Roberts:2010rn,Roberts:2011wy}.

Using Eqs.\,(\ref{njlgluon}), (\ref{RLvertex}), the gap equation becomes
\begin{equation}
 S^{-1}(p) =  i \gamma \cdot p + m +  \frac{16\pi}{3}\frac{\alpha_{\rm IR}}{m_G^2} \int\!\frac{d^4 q}{(2\pi)^4} \,
\gamma_{\mu} \, S(q) \, \gamma_{\mu}\,,   \label{gap-1}
\end{equation}
an equation in which the integral possesses a quadratic divergence, even in the chiral limit.  When the divergence is regularised in a Poincar\'e covariant manner, the solution is
\begin{equation}
\label{genS}
S(p)^{-1} = i \gamma\cdot p + M\,,
\end{equation}
where $M$ is momentum-independent and determined by
\begin{equation}
M = m + M\frac{4\alpha_{\rm IR}}{3\pi m_G^2} \int_0^\infty \!ds \, s\, \frac{1}{s+M^2}\,.
\end{equation}

Our regularisation procedure follows Ref.\,\cite{Ebert:1996vx}; i.e., we write
\begin{eqnarray}
\nonumber
\frac{1}{s+M^2} & = & \int_0^\infty d\tau\,{\rm e}^{-\tau (s+M^2)} \\
& \rightarrow & \int_{\tau_{\rm uv}^2}^{\tau_{\rm ir}^2} d\tau\,{\rm e}^{-\tau (s+M^2)}
\label{RegC}\\
& & =
\frac{{\rm e}^{- (s+M^2)\tau_{\rm uv}^2}-e^{-(s+M^2) \tau_{\rm ir}^2}}{s+M^2} \,, \label{ExplicitRS}
\end{eqnarray}
where $\tau_{\rm ir,uv}$ are, respectively, infrared and ultraviolet regulators.  It is apparent from Eq.\,(\ref{ExplicitRS}) that a finite value of $\tau_{\rm ir}=:1/\Lambda_{\rm ir}$ implements confinement by ensuring the absence of quark production thresholds \cite{Krein:1990sf,Chang:2011vu}.  Since Eq.\,(\ref{njlgluon}) does not define a renormalisable theory, then $\Lambda_{\rm uv}:=1/\tau_{\rm uv}$ cannot be removed but instead plays a dynamical role, setting the scale of all dimensioned quantities.  Using Eq.\,\eqref{RegC}, the gap equation becomes
\begin{equation}
M = m + M\frac{4\alpha_{\rm IR}}{3\pi m_G^2}\,\,{\cal C}^{\rm iu}(M^2)\,,
\label{gapactual}
\end{equation}
where ${\cal C}^{\rm iu}(M^2)/M^2 = \Gamma(-1,M^2 \tau_{\rm uv}^2) - \Gamma(-1,M^2 \tau_{\rm ir}^2)$, with $\Gamma(\alpha,y)$ being the incomplete gamma-function.

\subsection{Point-meson Bethe-Salpeter equation}
In rainbow-ladder truncation, with the interaction in Eq.\,(\ref{njlgluon}), the homogeneous Bethe-Salpeter equation for a colour-singlet meson is
\begin{equation}
\Gamma(k;P) =
-\frac{16 \pi}{3}\frac{\alpha_{\rm IR}}{m_G^2}
\int \! \frac{d^4q}{(2\pi)^4}\, \gamma_\mu \chi(q;P)\gamma_\mu \,,
\label{genbse}
\end{equation}
where $\chi(q;P) = S(q+P)\Gamma(q;P)S(q)$ and $\Gamma(q;P)$ is the meson's Bethe-Salpeter amplitude.  Since the integrand does not depend on the external relative-momentum, $k$, then a symmetry-preserving regularisation of Eq.\,(\ref{genbse}) yields solutions that are independent of $k$.  This is the defining characteristic of a pointlike composite particle.

With a dependence on the relative momentum forbidden by the interaction, then rainbow-ladder pseudoscalar and vector Bethe-Salpeter amplitudes take the form\footnote{We assume isospin symmetry throughout and hence do not include the Pauli isospin matrices explicitly.}
\begin{eqnarray}
\Gamma^\pi(P) &= &  i \gamma_5 E_\pi(P) + \frac{1}{M} \gamma_5 \gamma\cdot P F_\pi(P) \,,\\
\Gamma_\mu^\rho(P) & = & \gamma^T_\mu E_\rho(P), \label{rhobsa}
\end{eqnarray}
where $P_\mu \gamma^T_\mu = 0$ and $\gamma^T_\mu+\gamma^L_\mu=\gamma_\mu$.

Values of some meson-related quantities, of relevance herein and computed using the contact-interaction, are reported in Table~\ref{Table:static}.

\begin{table}[t]
\caption{Meson-related results obtained with $\alpha_{\rm IR}/\pi=0.93$ and (in GeV): $m=0.007$, $\Lambda_{\rm ir} = 0.24\,$, $\Lambda_{\rm uv}=0.905$ \protect\cite{Roberts:2011wy}.  The Bethe-Salpeter amplitudes are canonically normalised; $\kappa_\pi$ is the in-pion condensate \protect\cite{Brodsky:2010xf,Chang:2011mu}; and $f_{\pi,\rho}$ are the mesons' leptonic decay constants.  Empirical values are $\kappa_\pi \approx (0.22\,{\rm GeV})^3$ and \protect\cite{Nakamura:2010zzi} $f_\pi=0.092\,$GeV, $f_\rho=0.153\,$GeV.
\label{Table:static}
}
\begin{center}
\begin{tabular*}
{\hsize}
{
c@{\extracolsep{0ptplus1fil}}
c@{\extracolsep{0ptplus1fil}}
c@{\extracolsep{0ptplus1fil}}
|c@{\extracolsep{0ptplus1fil}}
c@{\extracolsep{0ptplus1fil}}
c@{\extracolsep{0ptplus1fil}}
c@{\extracolsep{0ptplus1fil}}
c@{\extracolsep{0ptplus1fil}}
c@{\extracolsep{0ptplus1fil}}}\hline
$E_\pi$ & $F_\pi$ & $E_\rho$ & $\kappa_\pi^{1/3}$ & $m_\pi$ & $m_\rho$ & $f_\pi$ & $f_\rho$ \\\hline
%
%
3.639 & 0.481 & 1.531 & 0.243 & 0.140 & 0.928 & 0.101 & 0.129\\\hline
\end{tabular*}
\end{center}
\end{table}

\subsection{Ward-Takahashi identities}
\label{sec:WTI}
No study of low-energy hadron observables is meaningful unless it ensures expressly that the vector and axial-vector Ward-Takahashi identities are satisfied.  Violation of these identities is a flaw of constituent-quark models that cannot be remedied.  The $m=0$ axial-vector identity states ($k_+ = k+P$)
\begin{equation}
\label{avwti}
P_\mu \Gamma_{5\mu}(k_+,k) = S^{-1}(k_+) i \gamma_5 + i \gamma_5 S^{-1}(k)\,,
\end{equation}
where $\Gamma_{5\mu}(k_+,k)$ is the axial-vector vertex, which is determined by
\begin{equation}
\Gamma_{5\mu}(k_+,k) =\gamma_5 \gamma_\mu
- \frac{16\pi}{3}\frac{\alpha_{\rm IR}}{m_G^2}
\int\frac{d^4q}{(2\pi)^4} \, \gamma_\alpha \chi_{5\mu}(q_+,q) \gamma_\alpha\,. \label{aveqn}
\end{equation}

One must implement a regularisation that maintains Eq.\,(\ref{avwti}).  That amounts to eliminating the quadratic and logarithmic divergences.  Their absence is just the circumstance under which a shift in integration variables is permitted, an operation required in order to prove Eq.\,(\ref{avwti}).  It is guaranteed so long as one implements the constraint \cite{GutierrezGuerrero:2010md,Roberts:2011cf,Roberts:2011wy}
\begin{equation}
0 = \int_0^1d\alpha \,
\left[ {\cal C}^{\rm iu}(\omega(M^2,\alpha,P^2))  + \, {\cal C}^{\rm iu}_1(\omega(M^2,\alpha,P^2))\right], \label{avwtiP}
\end{equation}
with
\begin{eqnarray}
\label{eq:omega}
\omega(M^2,\alpha,P^2) &=& M^2 + \alpha(1-\alpha) P^2\,,\\
\nonumber
{\cal C}^{\rm iu}_1(z) &=& - z (d/dz){\cal C}^{\rm iu}(z) \\
&= & z\left[ \Gamma(0,M^2 \tau_{\rm uv}^2)-\Gamma(0,M^2 \tau_{\rm ir}^2)\right] .\rule{2em}{0ex}
\label{eq:C1}
\end{eqnarray}

The vector Ward-Takahashi identity
\begin{equation}
\label{VWTI}
P_\mu i\Gamma^\gamma_\mu(k_+,k) = S^{-1}(k_+) - S^{-1}(k)\,,
\end{equation}
wherein $\Gamma^\gamma_\mu$ is the dressed-quark-photon vertex, is crucial for a sensible study of a bound-state's electromagnetic form factors \cite{Roberts:1994hh}.  The vertex must be dressed at a level consistent with the truncation used to compute the bound-state's Bethe-Salpeter or Faddeev amplitude.  Herein this means the vertex should be determined from the following inhomogeneous Bethe-Salpeter equation:
\begin{equation}
\label{GammaQeq}
\Gamma_\mu(Q) = \gamma_\mu -
\frac{16\pi}{3} \frac{\alpha_{\rm IR}}{m_G^2}
\int \frac{d^4 q}{(2\pi)^4} \, \gamma_\alpha \chi_\mu(q_+,q) \gamma_\alpha\,,
\end{equation}
where $\chi_\mu(q_+,q) = S(q+P) \Gamma_\mu (Q) S(q)$.  Owing to the momentum-independent nature of the interaction kernel, the general form of the solution is
\begin{equation}
\label{GammaQ}
\Gamma_\mu(Q) = \gamma^T_\mu P_T(Q^2) + \gamma_\mu^L P_L(Q^2)\,.
\end{equation}

Inserting Eq.\,(\ref{GammaQ}) into Eq.\,(\ref{GammaQeq}), one readily obtains
\begin{equation}
\label{PL0}
P_L(Q^2)= 1\,,
\end{equation}
owing to corollaries of Eq.\,(\ref{avwti}).  Using these same identities, one finds \cite{Roberts:2011wy}
\begin{equation}
\label{PTQ2}
P_T(Q^2)= \frac{1}{1+K_\gamma(Q^2)},
\end{equation}
with ($\overline{\cal C}_1(z) = {\cal C}_1(z)/z$)
\begin{eqnarray}
\nonumber
\lefteqn{K_\gamma(Q^2) = \frac{4\alpha_{\rm IR}}{3\pi m_G^2}}\\
%
& & \times \int_0^1d\alpha\, \alpha(1-\alpha) Q^2\,  \overline{\cal C}^{iu}_1(\omega(M^2,\alpha,Q^2))\,. \label{Kgamma}
\end{eqnarray}

\subsection{Diquark Bethe-Salpeter amplitudes}
\label{qqBSA}
In the rainbow-ladder truncation, colour-antitriplet quark-quark correlations (diquarks) are described by an homogeneous Bethe-Salpeter equation that is readily inferred from Eq.\,(\ref{genbse}); viz., following Ref.\,\cite{Cahill:1987qr} and expressing the diquark amplitude as
\begin{equation}
\Gamma^c_{qq}(k;P) = \Gamma_{qq}(k;P) C^\dagger H^{c}
\end{equation}
then
\begin{equation}
\Gamma_{qq}(k;P) = -\frac{8 \pi }{3}\frac{\alpha_{\rm IR}}{m_G^2} \int \! \frac{d^4q}{(2\pi)^4}\, \gamma_\mu \chi_{qq}(q;P)\gamma_\mu \,.
\label{genbseqq}
\end{equation}
Hence, one may obtain the mass and amplitude for a diquark with spin-parity $J^P$ from the equation for a $J^{-P}$-meson in which the only change is a halving of the interaction strength.  The flipping of the sign in parity occurs because fermions and antifermions have opposite parity.

Scalar and axial-vector quark-quark correlations are dominant in studies of the nucleon and Roper:
\begin{eqnarray}
\label{scqqbsa}
\Gamma^{0^+}_{qq}(P) &= &  i \gamma_5 E_{qq 0^+}(P) + \frac{1}{M} \gamma_5 \gamma\cdot P F_{qq 0^+}(P) \,,\rule{2em}{0ex}\\
\Gamma_{qq\, \mu}^{1^+}(P) & = & \gamma^T_\mu E_{qq 1^+}(P). \label{avqqbsa}
\end{eqnarray}
These amplitudes are canonically normalised:
\begin{equation}
P_\mu = 2 {\rm tr}\!\!\int\frac{d^4q}{(2\pi)^4} \Gamma_{qq}^{0^+}(-P) \frac{\partial}{\partial_\mu} S(q+P) \Gamma_{qq}^{0^+}(P) S(q);
\end{equation}
and
\begin{equation}
P_\mu = \frac{2}{3} {\rm tr}\!\!\int\frac{d^4q}{(2\pi)^4} \Gamma_{qq\,\alpha}^{1^+}(-P) \frac{\partial}{\partial_\mu} S(q+P) \Gamma_{qq\,\alpha}^{1^+}(P) S(q).
\end{equation}

\section{Faddeev Equation}
\label{sec:Faddeev}
We describe the dressed-quark-cores of the nucleon and Roper via solutions of a Poincar\'e-covariant Faddeev equation \cite{Cahill:1988dx}.  The equation is derived following upon the observation that an interaction which describes mesons also generates diquark correlations in the colour-$\bar 3$ channel \cite{Cahill:1987qr}.  The fidelity of the diquark approximation to the quark-quark scattering kernel is verified by recent studies \cite{Eichmann:2011vu}.

Within this approach, a $J=\frac{1}{2}$ baryon is represented by a Faddeev amplitude
\begin{equation}
\label{PsiNucleon}
\Psi = \Psi_1 + \Psi_2 + \Psi_3  \,,
\end{equation}
where the subscript identifies the bystander quark and, e.g., $\Psi_{1,2}$ are obtained from $\Psi_3$ by a cyclic permutation of all the quark labels.  We employ a simple but realistic representation of $\Psi$.  The spin- and isospin-$\frac{1}{2}$ nucleon and Roper are each a sum of scalar and axial-vector diquark correlations:
\begin{equation}
\label{Psi} \Psi_3(p_i,\alpha_i,\tau_i) = {\cal N}_3^{0^+} + {\cal N}_3^{1^+},
\end{equation}
with $(p_i,\alpha_i,\tau_i)$ the momentum, spin and isospin labels of the
quarks constituting the bound state, and $P=p_1+p_2+p_3$ the system's total momentum.

The scalar diquark piece in Eq.\,(\ref{Psi}) is
\begin{eqnarray}
\nonumber
{\cal N}_3^{0^+}(p_i,\alpha_i,\tau_i) &= &[\Gamma^{0^+}(\frac{1}{2}p_{[12]};K)]_{\alpha_1
\alpha_2}^{\tau_1 \tau_2}\\
&&  \rule{0em}{0ex} \times \Delta^{0^+}(K) \,[{\cal S}(\ell;P) u(P)]_{\alpha_3}^{\tau_3},
\label{calS}
\end{eqnarray}
where: the spinor satisfies Eq.\,(\ref{DiracEqn}), with $M$ the mass obtained by solving the Faddeev equation, and it is also a spinor in isospin space with $\varphi_+= {\rm col}(1,0)$ for the charge-one state and $\varphi_-= {\rm col}(0,1)$ for the neutral state; $K= p_1+p_2=: p_{\{12\}}$, $p_{[12]}= p_1 - p_2$, $\ell := (-p_{\{12\}} + 2 p_3)/3$;
\begin{equation}
\label{scalarqqprop}
\Delta^{0^+}(K) = \frac{1}{K^2+m_{qq_{0^+}}^2}
\end{equation}
is a propagator for the scalar diquark formed from quarks $1$ and $2$, with $m_{0^+}$ the mass-scale associated with this correlation, and $\Gamma^{0^+}\!$ is the canonically-normalised Bethe-Salpeter amplitude describing their relative momentum correlation, Sec.\,\ref{qqBSA}; and ${\cal S}$, a $4\times 4$ Dirac matrix, describes the relative quark-diquark momentum correlation.  The colour antisymmetry of $\Psi_3$ is implicit in $\Gamma^{J^P}\!\!$, with the Levi-Civita tensor, $\epsilon_{c_1 c_2 c_3}$, expressed via the antisymmetric Gell-Mann matrices; viz., defining
\begin{eqnarray}
\label{Hmatrices}
&& \{H^1=i\lambda^7,H^2=-i\lambda^5,H^3=i\lambda^2\}\,,\\
&\mbox{then}&  \epsilon_{c_1 c_2 c_3}= (H^{c_3})_{c_1 c_2}.
\end{eqnarray}

The axial-vector component in Eq.\,(\ref{Psi}) is
\begin{eqnarray}
\nonumber
{\cal N}^{1^+}(p_i,\alpha_i,\tau_i) & =&  [{\tt t}^i\,\Gamma_\mu^{1^+}(\frac{1}{2}p_{[12]};K)]_{\alpha_1
\alpha_2}^{\tau_1 \tau_2}\\
&& \times \Delta_{\mu\nu}^{1^+}(K)\,
[{\cal A}^{i}_\nu(\ell;P) u(P)]_{\alpha_3}^{\tau_3}\,,
\label{calA}
\end{eqnarray}
where the symmetric isospin-triplet matrices are
\begin{equation}
{\tt t}^+ = \frac{1}{\surd 2}(\tau^0+\tau^3) \,,\;
{\tt t}^0 = \tau^1\,,\;
{\tt t}^- = \frac{1}{\surd 2}(\tau^0-\tau^3)\,,
\end{equation}
and the other elements in Eq.\,(\ref{calA}) are straightforward generalisations of those in Eq.\,(\ref{calS}) with, e.g.,
\begin{equation}
\label{avqqprop}
\Delta_{\mu\nu}^{1^+}(K) = \frac{1}{K^2+m_{qq_{1^+}}^2} \, \left(\delta_{\mu\nu} + \frac{K_\mu K_\nu}{m_{qq_{1^+}}^2}\right) \,.
\end{equation}

One can now write the Faddeev equation for $\Psi_3$:
\begin{eqnarray}
\nonumber
\lefteqn{
 \left[ \begin{array}{r}
{\cal S}(k;P)\, u(P)\\
{\cal A}^i_\mu(k;P)\, u(P)
\end{array}\right]}\\
& =&  -\,4\,\int\frac{d^4\ell}{(2\pi)^4}\,{\cal M}(k,\ell;P)
\left[
\begin{array}{r}
{\cal S}(\ell;P)\, u(P)\\
{\cal A}^j_\nu(\ell;P)\, u(P)
\end{array}\right] .\rule{1em}{0ex}
\label{FEone}
\end{eqnarray}
The kernel in Eq.\,(\ref{FEone}) is
\begin{equation}
\label{calM} {\cal M}(k,\ell;P) = \left[\begin{array}{cc}
{\cal M}_{00} & ({\cal M}_{01})^j_\nu \\
({\cal M}_{10})^i_\mu & ({\cal M}_{11})^{ij}_{\mu\nu}\rule{0mm}{3ex}
\end{array}
\right] ,
\end{equation}
with
\begin{eqnarray}
\nonumber
 {\cal M}_{00} &=& \Gamma^{0^+}\!(k_q-\ell_{qq}/2;\ell_{qq})\,
S^{\rm T}(\ell_{qq}-k_q) \\
&& \times \,\bar\Gamma^{0^+}\!(\ell_q-k_{qq}/2;-k_{qq})\,
S(\ell_q)\,\Delta^{0^+}(\ell_{qq}) \,, \rule{2em}{0ex}
\end{eqnarray}
where: $\ell_q=\ell$, $k_q=k$, $\ell_{qq}=-\ell+ P$,
$k_{qq}=-k+P$ and the superscript ``T'' denotes matrix transpose; and
\begin{eqnarray}
\nonumber
({\cal M}_{01})^j_\nu &=& {\tt t}^j \,
\Gamma_\mu^{1^+}\!(k_q-\ell_{qq}/2;\ell_{qq}) S^{\rm T}(\ell_{qq}-k_q)\,\\
&& \rule{-1.5em}{0ex} \times \bar\Gamma^{0^+}\!(\ell_q-k_{qq}/2;-k_{qq})\,
S(\ell_q)\,\Delta^{1^+}_{\mu\nu}(\ell_{qq}) , \rule{2.2em}{0ex} \label{calM01} \\
\nonumber
({\cal M}_{10})^i_\mu &=& \Gamma^{0^+}\!(k_q-\ell_{qq}/2;\ell_{qq})\,
S^{\rm T}(\ell_{qq}-k_q)\,{\tt t}^i\, \\
&& \rule{-1.5em}{0ex}\times \bar\Gamma_\mu^{1^+}\!(\ell_q-k_{qq}/2;-k_{qq})\,
S(\ell_q)\,\Delta^{0^+}(\ell_{qq}) , \rule{2.2em}{0ex}\\
\nonumber
({\cal M}_{11})^{ij}_{\mu\nu} &=& {\tt t}^j\,
\Gamma_\rho^{1^+}\!(k_q-\ell_{qq}/2;\ell_{qq})\, S^{\rm T}(\ell_{qq}-k_q)\,{\tt t}^i\,\\
&& \rule{-1.5em}{0ex}\times  \bar\Gamma^{1^+}_\mu\!(\ell_q-k_{qq}/2;-k_{qq})\,
S(\ell_q)\,\Delta^{1^+}_{\rho\nu}(\ell_{qq}) . \rule{2.2em}{0ex}\label{calM11}
\end{eqnarray}

Our dressed-quark propagator is described in Sec.\,\ref{sec:gap} and the diquark propagators are given in Eqs.\,(\ref{scalarqqprop}), (\ref{avqqprop}), so the Faddeev equation is complete once the diquark Bethe-Salpeter amplitudes are known.  They are reviewed in Sec.\,\ref{qqBSA}.  We note here, however, that we follow Ref.\,\cite{Roberts:2011cf} and employ a simplification of the kernel; viz., in the Faddeev equation, the quark exchanged between the diquarks is represented as
\begin{equation}
S^{\rm T}(k) \to \frac{g_N^2}{M}\,,
\label{staticexchange}
\end{equation}
where $g_N=1.18$ \cite{Roberts:2011cf}.  This is a variant of the so-called ``static approximation,'' which itself was introduced in Ref.\,\cite{Buck:1992wz} and has subsequently been used in studying a range of nucleon properties \cite{Bentz:2007zs}.  In combination with diquark correlations generated by Eq.\,(\ref{njlgluon}), whose Bethe-Salpeter amplitudes are momentum-independent, Eq.\,(\ref{staticexchange}) generates Faddeev equation kernels which themselves are momentum-independent.  The dramatic simplifications which this produces are the merit of Eq.\,(\ref{staticexchange}).

The general forms of the matrices ${\cal S}(\ell;P)$ and ${\cal A}^i_\nu(\ell;P)$, which describe the momentum-space correlation between the quark and diquark in the nucleon and Roper, are described in Refs.\,\cite{Oettel:1998bk,Cloet:2007pi}.  However, with the interaction described in Sec.\,\ref{sec:gap} augmented by Eq.\,(\ref{staticexchange}), they simplify greatly; viz.,
\begin{subequations}
\label{FaddeevAmp}
\begin{eqnarray}
{\cal S}(P) &=& s(P) \,\mbox{\boldmath $I$}_{\rm D}\,,\\
{\cal A}^i_\mu(P) &=& a_1^i(P) \gamma_5\gamma_\mu + a_2^i(P) \gamma_5 \hat P_\mu \,,i=+,0\,,\rule{2em}{0ex}
\end{eqnarray}
\end{subequations}
with the scalars $s$, $a_{1,2}^i$ independent of the relative quark-diquark momentum and $\hat P^2=-1$.

The mass of the ground-state nucleon is then determined by a $5\times 5$ matrix Faddeev equation; viz., $\Psi = K \Psi$, with eigenvector
\begin{equation}
\Psi(P) = \left[\begin{array}{c}
s(P)\\[0.7ex]
a_1^+(P)\\[0.7ex]
a_1^0(P)\\[0.7ex]
a_2^+(P)\\[0.7ex]
a_2^0(P)\end{array}\right],
\end{equation}
and kernel
\begin{widetext}
\begin{equation}
K(P) =
\left[ \begin{array}{ccccc}
K^{00}_{ss} & -\surd 2 \, K^{01}_{sa_1} & K^{01}_{sa_1} & -\surd 2\, K^{01}_{sa_2} & K^{01}_{sa_2}\\[0.7ex]
-\surd 2\, K^{10}_{a_1 s} & 0 & \surd 2\, K^{11}_{a_1 a_1} & 0 & \surd 2\, K^{11}_{a_1 a_2}\\[0.7ex]
K^{10}_{a_1 s} & \surd 2 \, K^{11}_{a_1 a_1} & K^{11}_{a_1 a_1} & \surd 2\,K^{11}_{a_1 a_2} & K^{11}_{a_1 a_2} \\[0.7ex]
-\surd 2\, K^{10}_{a_2 s} & 0 & \surd 2\, K^{11}_{a_2 a_1} & 0 & \surd 2\, K^{11}_{a_2 a_2} \\[0.7ex]
K^{10}_{a_2 s} & \surd 2\, K^{11}_{a_2 a_1} & K^{11}_{a_2 a_1} & \surd 2\, K^{11}_{a_2 a_2} & K^{11}_{a_2 a_2}
\end{array}
\right],
\end{equation}
%
%
%
%
%
constructed using: $c_N = g_N^2/(4 \pi^2 M)$;
%
\begin{equation}
\sigma_N^0  = \sigma_N(\alpha,M,m_{qq_{0^+}},m_N) :=
(1-\alpha)\,M^2 + \alpha\,m_{qq_{0^+}}^2 - \alpha (1-\alpha) m_N^2\,,\rule{1em}{0ex}
\sigma_N^1 = \sigma_N(\alpha,M,m_{qq_{1^+}},m_N) \,;
\end{equation}
and
\begin{subequations}
\begin{eqnarray}
K^{00}_{ss} & = & K^{00}_{EE}+K^{00}_{EF}+K^{00}_{FF}\,,\\
K^{00}_{EE} & = & c_N E_{qq_{0^+}}^2 \!
\int_0^1 d\alpha \,\overline{\cal C}^{\rm iu}_1(\sigma_N^0)
(\alpha m_N + M)\,,\\
K^{00}_{EF} & = & - 2 c_N E_{qq_{0^+}} F_{qq_{0^+}}\frac{m_N}{M} \!
\int_0^1 d\alpha \,\overline{\cal C}^{\rm iu}_1(\sigma_N^0)
(1-\alpha) (\alpha m_N + M)\,,\\
K^{00}_{FF} & = & c_N F_{qq_{0^+}}^2\frac{m_{qq_{0^+}}^2}{M^2} \!
\int_0^1 d\alpha \,\overline{\cal C}^{\rm iu}_1(\sigma_N^0)(\alpha m_N + M)\,;\\
K^{01}_{s a_1} & = & K^{01}_{s_E a_1} + K^{01}_{s_F a_1}\,,\\
K^{01}_{s_E a_1} &=& c_N \frac{E_{qq_{0^+}}E_{qq_{1^+}}}{m_{qq_{1^+}}^2}\!
\int_0^1 d\alpha \,\overline{\cal C}^{\rm iu}_1(\sigma_N^1)
( m_{qq_{1^+}}^2 (3 M + \alpha m_N) + 2 \alpha (1-\alpha)^2 m_N^3 )\,,
\rule{5em}{0ex}
\end{eqnarray}
\begin{eqnarray}
K^{01}_{s_F a_1} &=& -c_N \frac{F_{qq_{0^+}}E_{qq_{1^+}}}{m_{qq_{1^+}}^2} \frac{m_N}{M}\!
\int_0^1 d\alpha \,\overline{\cal C}^{\rm iu}_1(\sigma_N^1)
(1-\alpha)
(m_{qq_{1^+}}^2 (M + 3 \alpha m_N) + 2 (1-\alpha)^2 M m_N^2) \,;\\
K^{01}_{s a_2} & = & K^{01}_{s_E a_2} + K^{01}_{s_F a_2}\,,\\
K^{01}_{s_E a_2} & = & c_N \frac{E_{qq_{0^+}}E_{qq_{1^+}}}{m_{qq_{1^+}}^2}\!
\int_0^1 d\alpha \,\overline{\cal C}^{\rm iu}_1(\sigma_N^1)
(\alpha m_N - M) ((1-\alpha)^2 m_N^2-m_{qq_{1^+}}^2)\,,\\
K^{01}_{s_F a_2} & = & c_N \frac{F_{qq_{0^+}}E_{qq_{1^+}}}{m_{qq_{1^+}}^2} \frac{m_N}{M} \!
\int_0^1 d\alpha \,\overline{\cal C}^{\rm iu}_1(\sigma_N^1)
(1-\alpha)(\alpha m_N - M) ((1-\alpha)^2 m_N^2-m_{qq_{1^+}}^2)\,;\\
K^{10}_{a_1 s} & = & K^{10}_{a_1 s_E} + K^{10}_{a_1 s_F}\,,\\
K^{10}_{a_1 s_E} & = & \frac{c_N}{3}\frac{E_{qq_{0^+}}E_{qq_{1^+}}}{m_{qq_{1^+}}^2} \!
\int_0^1 d\alpha \,\overline{\cal C}^{\rm iu}_1(\sigma_N^0)
(\alpha m_N + M) (2 m_{qq_{1^+}}^2 + (1-\alpha)^2 m_N^2)\,,\\
K^{10}_{a_1 s_F} & = & -\frac{c_N}{3}\frac{F_{qq_{0^+}}E_{qq_{1^+}}}{m_{qq_{1^+}}^2} \frac{m_N}{M} \!
\int_0^1 d\alpha \,\overline{\cal C}^{\rm iu}_1(\sigma_N^0)
(1-\alpha)(2 m_{qq_{1^+}}^2 + (1-\alpha)^2 m_N^2) (\alpha m_N + M)\,;\\
K^{10}_{a_2 s} & = & K^{10}_{a_2 s_E} + K^{10}_{a_2 s_F}\,,\\
K^{10}_{a_2 s_E} & = & \frac{c_N}{3} \frac{E_{qq_{0^+}}E_{qq_{1^+}}}{m_{qq_{1^+}}^2} \!
\int_0^1 d\alpha \,\overline{\cal C}^{\rm iu}_1(\sigma_N^0)
(\alpha m_N + M) (m_{qq_{1^+}}^2 - 4 (1-\alpha)^2 m_N^2),\\
K^{10}_{a_2 s_F} & = & \frac{c_N}{3} \frac{F_{qq_{0^+}}E_{qq_{1^+}}}{m_{qq_{1^+}}^2}\frac{m_N}{M} \!
\int_0^1 d\alpha \,\overline{\cal C}^{\rm iu}_1(\sigma_N^0)
(1-\alpha) (5 m_{qq_{1^+}}^2-2(1-\alpha)^2 m_N^2)(\alpha m_N + M)\,;\\
K^{11}_{a_1 a_1} & = & -\frac{c_N}{3}\frac{E_{qq_{1^+}}^2}{m_{qq_{1^+}}^2} \!
\int_0^1 d\alpha \,\overline{\cal C}^{\rm iu}_1(\sigma_N^1)
[ 2 m_{qq_{1^+}}^2 (M-\alpha m_N) + (1-\alpha)^2 m_N^2 (M+5 \alpha m_N)]\,;\\
K^{11}_{a_1 a_2} & = & -\frac{2 c_N}{3}\frac{E_{qq_{1^+}}^2}{m_{qq_{1^+}}^2} \!
\int_0^1 d\alpha \,\overline{\cal C}^{\rm iu}_1(\sigma_N^1)
(-m_{qq_{1^+}}^2+(1-\alpha)^2 m_N^2) (\alpha m_N - M)\,;\\
K^{11}_{a_2 a_1} & = & -\frac{c_N}{3}\frac{E_{qq_{1^+}}^2}{m_{qq_{1^+}}^2} \!
\int_0^1 d\alpha \,\overline{\cal C}^{\rm iu}_1(\sigma_N^1)
[m_{qq_{1^+}}^2(11 \alpha  m_N + M) - 2(1-\alpha)^2 m_N^2 (7\alpha m_N + 2 M)]\,;\\
K^{11}_{a_2 a_2}  & = & - \frac{5 c_N}{3} \frac{E_{qq_{1^+}}^2}{m_{qq_{1^+}^2}} \!
\int_0^1 d\alpha \,\overline{\cal C}^{\rm iu}_1(\sigma_N^1)
(m_{qq_{1^+}}^2 - (1-\alpha)^2 m_N^2) (\alpha m_N - M)\,.
\end{eqnarray}
\end{subequations}
\end{widetext}
The computation of this kernel is detailed in Ref.\,\cite{Roberts:2011cf}.  The eigenvectors exhibit the pattern:
\begin{equation}
\label{aia0}
a_i^+ = -\sqrt{ 2} a_i^0,\; i=1,2.
\end{equation}

The kernel for the Roper resonance has the same form but there is one change; namely, the functions ${\cal C}^{\rm iu}$ are replaced by functions ${\cal F}^{\rm iu} = {\cal C}^{\rm iu} - d_{\cal F} {\cal D}^{\rm iu}$ where
\begin{eqnarray}
\nonumber
\lefteqn{{\cal D}^{\rm iu}(\omega(M^2,\alpha,P^2))= \int_0^\infty ds\,s\,\frac{s}{s+\omega }\rule{2em}{0ex}}\\
&\to&  \int_{\tau_{\rm uv}^2}^{\tau_{\rm iu}^2} d\tau\, \frac{2}{\tau^3} \,
\exp\left[-\tau \omega(M^2,\alpha,P^2)\right], \rule{1em}{0ex}
\end{eqnarray}
${\cal F}^{\rm iu}_1(z) = - z (d/dz){\cal F}^{\rm iu}(z)$ and $\overline{\cal F}_1(z) = {\cal F}_1(z)/z$.  As explained in Sec.\,3.2 of Ref.\,\cite{Roberts:2011cf}, this has the effect of inserting a zero at $q^2=1/d_{\cal F}$ in the amplitude for the nucleon's excitation, which then has the structure of a radial excitation of the bystander quark with respect to the diquark ``core.''

Solving for the Roper with this kernel and $M d_{\cal F}^{1/2}=0.88$ we obtain
\begin{equation}
\label{FaddeevAmpR}
\begin{array}{lccccc}
\mbox{mass~(GeV)} & s & a_1^+ & a_1^0 & a_2^+ & a_2^0 \\
m_R = 1.72  & -0.0828 & ~~0.590 & -0.417 & -0.561~ & 0.397~
\end{array}.
\end{equation}
The eigenvector differs from that listed in Table~\ref{Table:FE}B for reasons that are explained in App.\,\ref{App:transitioncurrent}.

\section{Electromagnetic Current}
\label{App:current}
Using the properties of our baryon spinors, the current in Eq.\,(\ref{Belastic}) can be rewritten in the form
\begin{eqnarray}
\nonumber
{\cal J}^B_\mu(Q) &=& i e \Lambda^B_+(P_f) \left[ \gamma_\mu \, F_{1B}(Q^2) \right.\\
&& \left. + \frac{1}{2M_B} \sigma_{\mu\nu} Q_\nu F_{2B}(Q^2)\right] \Lambda^B_+(P_i),
\label{CalJB}
\end{eqnarray}
where the positive-energy projection operator is defined in Eq.\,(\ref{Lplus}).  In this connection each of the three diagrams in Fig.\,\ref{fig:current} can similarly be expressed
\begin{equation}
\nonumber
L_\mu^k(Q)
= \Lambda^B_+(P_f) {\cal I}^k_\mu(P_f,P_i)\Lambda^B_+(P_i), \; k=1,2,3.
\end{equation}
In being explicit, we will focus on the elastic form factors for the charged baryon.  N.B.\ For the neutral particles, one simply exchanges the flavours of the doubly- and singly-represented quarks.

\subsection{Diagram 1}
\label{sec:D1}
The uppermost diagram in Fig.\,\ref{fig:current} describes a photon coupling directly to a dressed-quark, through the vertex described in App.\,\ref{sec:WTI}.  It can be seen to represent the following three expressions, the first involving the scalar diquark and the second two, the axial-vector diquarks:
\begin{equation}
\label{CalI1s}
{\cal I}^1_{s\mu} = s(P_i)^2 \!\int_\ell S(\ell_f) i e_u\gamma^T_\mu P_T(Q^2) S(\ell_i)
\Delta^{0^+}(-\ell),
\end{equation}
where $\int_\ell = \int\frac{d^4\ell}{(2\pi)^4}$, $\ell_{\pm (i,f)} = \ell\pm P_{i,f}$, $e_u=2/3$; and
\begin{eqnarray}
\nonumber {\cal I}^{1+}_{j\mu} &= &  a_{j}^+(P)^2 \!\!
\int_\ell \bar M_{j\alpha} S(\ell_f) \\
&& \times i e_d\gamma^T_\mu P_T(Q^2) S(\ell_i) M_{j\beta}
\Delta_{\alpha\beta}^{1^+}(-\ell), \\ 
\nonumber {\cal I}^{10}_{j\mu} &= &  a_{j}^0(P)^2 \!\!
\int_\ell M_{j\alpha} S(\ell_f) \\
&& \times i e_u\gamma^T_\mu P_T(Q^2) S(\ell_i) M_{j\beta}
\Delta_{\alpha\beta}^{1^+}(-\ell), 
\end{eqnarray}
with $e_d=-1/3$, $j=1,2$ and $M_{1\alpha} = \gamma_5 \gamma_\beta $, $M_{2\alpha} =  \gamma_5 \hat P_\mu$.  If one assumes isospin symmetry, as herein, then it is notable that owing to Eq.\,(\ref{aia0})
\begin{equation}
\label{isospinFF}
{\cal I}^{1+}_{j\mu} + {\cal I}^{10}_{j\mu} \equiv 0\,,\; j=1,2\,,
\end{equation}
which means diagrams with axial-vector diquark spectators do not contribute to charged-particle form factors.

\subsection{Diagram 2}
The second diagram in Fig.\,\ref{fig:current} depicts the photon scattering elastically from a diquark, with the dressed-quark as spectator.  Again, it can be expressed through the sum of three separate terms, the first involving the scalar diquark:
\begin{eqnarray}
\nonumber
{\cal I}^2_{s\mu} &= & s(P)^2 \! \int_\ell S(\ell) \Delta^{0^+}(-\ell_{-f})  \\
&& \times e_{[ud]}\Gamma_\mu^{0+}(-\ell_{-f},-\ell_{-i}) \Delta^{0^+}(-\ell_{-i}),
\label{CalI2s}
\end{eqnarray}
where $e_{[ud]}=1/3$.  Here $\Gamma_\mu^{0+}$ is the dressed-photon--scalar-diquark vertex, computed in Ref.\,\cite{Roberts:2011wy}:
\begin{equation}
\Gamma_\mu^{0+}(-\ell_{-f},-\ell_{-i}) = -(\ell_{-f}+\ell_{-i}) F_{0^+}(Q^2)\,,\\
\end{equation}
with the following expression providing an accurate interpolation on the domain $Q^2\in[-m_\rho^2,10]\,$GeV$^2$, $m_\rho$ is the $\rho$-meson's mass,
\begin{eqnarray}
F_{0^+}(Q^2) &\stackrel{\rm interpolation}{=} &
\frac{1+0.25\,Q^2+0.027\,Q^4}{1+1.27 \, Q^2 + 0.13 \,Q^4}\,. \rule{1em}{0ex}
\end{eqnarray}

The remaining terms involve elastic scattering from the axial-vector diquark:
\begin{eqnarray}
\nonumber
{\cal I}^{2+}_{j\mu} & = & a_{j}^+(P)^2 \! \int_\ell \bar M_{j\alpha} S(\ell) \Delta_{\alpha\rho}^{1^+}(-\ell_f)  \\
& & \times e_{\{uu\}} \Gamma_{\mu,\rho\sigma}^{1+}(-\ell_{-f},-\ell_{-i}) \Delta_{\sigma\beta}^{1^+}(-\ell_{-i})M_{j\beta},\\
\nonumber
{\cal I}^{20}_{j\mu} & = & a_{j}^0(P)^2 \! \int_\ell \bar M_{j\alpha} S(\ell) \Delta_{\alpha\rho}^{1^+}(-\ell_f)  \\
& & \times e_{\{ud\}} \Gamma_{\mu,\rho\sigma}^{1+}(-\ell_{-f},-\ell_{-i}) \Delta_{\sigma\beta}^{1^+}(-\ell_{-i})M_{j\beta},\rule{2em}{0ex}
\end{eqnarray}
with $e_{\{uu\}}=4/3$, $e_{\{ud\}}=1/3$ and
\begin{eqnarray}
\nonumber \Gamma_{\mu,\rho\sigma}^{1+}(k_f=K+Q/2,k_i=K-Q/2) \\
= \sum_{j=1}^3 T_{\mu,\rho\sigma}^j(K,Q) \, F_j^{1+}(Q^2)\,,
\end{eqnarray}
where
\begin{subequations}
\begin{eqnarray}
T_{\mu,\rho\sigma}^1(K,Q) & = & 2 K_\mu\, {\cal P}^T_{\rho\alpha}(p^i) \, {\cal P}^T_{\alpha\sigma}(p^f)\,,\\
\nonumber
T_{\mu,\rho\sigma}^2(K,Q) & = & \left[Q_\rho - p^i_\rho \frac{Q^2}{2 m_{1^+}^2}\right] {\cal P}^T_{\mu\sigma}(p^f) \\
&& - \left[Q_\sigma + p^f_\sigma \frac{Q^2}{2 m_{1^+}^2}\right] {\cal P}^T_{\mu\rho}(p^i)\,, \\
\nonumber
T_{\mu,\rho\sigma}^3(K,Q) & = & \frac{K_\mu}{m_{1^+}^2}\! \left[Q_\rho - p^i_\rho \frac{Q^2}{2 m_{1^+}^2}\right]\!\! \left[Q_\sigma + p^f_\sigma \frac{Q^2}{2 m_{1^+}^2}\right] \,,\\
&&
\end{eqnarray}
\end{subequations}
${\cal P}^T_{\rho\sigma}(p) = \delta_{\rho\sigma} - p_\rho p_\sigma/p^2$.  The electric, magnetic and quadrupole form factors of the axial-vector diquark are constructed as follows:
\begin{subequations}
\label{GsFs}
\begin{eqnarray}
G_E^{1+}(Q^2) & = & F_1^{1+}(Q^2)+\frac{2}{3} \eta G^{1+}_Q(Q^2)\,,\\
G_M^{1+}(Q^2) & = & - F^{1+}_2(Q^2)\,,\\
\nonumber
G_Q^{1+}(Q^2) & = & F^{1+}_1(Q^2) \\
&& + F^{1+}_2(Q^2) + \left[1+\eta\right] F^{1+}_3(Q^2)\,,\rule{1em}{0ex}
\end{eqnarray}
\end{subequations}
where $\eta=Q^2/[4 m_{1^+}^2]$.  These quantities were computed in Ref.\,\cite{Roberts:2011wy} and the following functions provide accurate interpolations on $Q^2\in[-m_\rho^2,10]\,$GeV$^2$:
\begin{subequations}
\label{GsAVqq}
\begin{eqnarray}
G_E^{1^+}(Q^2) &\stackrel{\rm interpolation}{=} &
\frac{1-0.16\,Q^2}{1+1.17 \, Q^2 + 0.012 \,Q^4}\,,\\
G_M^{1^+}(Q^2) &\stackrel{\rm interpolation}{=}&
\frac{2.13-0.19\,Q^2}{1+1.07 \, Q^2 - 0.10 \,Q^4}\,,\\
G_Q^{1^+}(Q^2) &\stackrel{\rm interpolation}{=}&
-\frac{0.81-0.029\,Q^2}{1+1.11 \, Q^2 - 0.054 \,Q^4}\,.\rule{2em}{0ex}
\end{eqnarray}
\end{subequations}

\subsection{Diagram 3}
\label{sec:D3}
The last diagram depicts a dressed-quark spectator to a photon induced transition between scalar and axial-vector diquarks.  It may be constructed from a sum
\begin{equation}
\label{calI3j}
{\cal I}^{3}_{j\mu} = {\cal I}^{301}_{j\mu} +{\cal I}^{310}_{j\mu}
\end{equation}
where
\begin{eqnarray}
\nonumber
{\cal I}^{301}_{j\mu} & = & s(P)a_{j}^0(P)\! \int_\ell S(\ell)
\Delta^{0^+}(-\ell_f)  \\
& & \times ie_{\{ud\}} \Gamma_{\mu \rho}^{01}(Q,-\ell_{-i}) \Delta_{\rho\beta}^{1^+}(-\ell_i)M_{j\beta},\rule{1em}{0ex}\\
\nonumber
{\cal I}^{310}_{j\mu} & = & s(P)a_{j}^0(P)\! \int_\ell \bar M_{j\alpha} S(\ell)
\Delta_{\alpha \rho}^{1^+}(-\ell_f)  \\
& & \times ie_{\{ud\}} \Gamma_{\rho\mu}^{10}(-\ell_{-f},Q) \Delta^{0^+}(-\ell_i),\rule{1em}{0ex}
\end{eqnarray}
with
\begin{equation}
\Gamma_{\rho\mu}^{10}(k_2,k_1) = \Gamma_{\rho\mu}^{01}(-k_2,k_1) = \Gamma_{\mu\rho}^{01}(k_1,k_2)
\end{equation}
and
\begin{eqnarray}
\Gamma_{\mu \rho}^{01}(k_1,k_2) &=&
\frac{g_{01}}{m_{qq_{1^+}}} \, \epsilon_{\mu\rho\alpha\beta}k_{1\alpha} k_{2\beta}\, G^{0 1}(Q^2).
\end{eqnarray}

The coupling and form factor were computed in Ref.\,\cite{Roberts:2011wy}, with the results: $g_{01}=0.78$; and a function for which an accurate interpolation on $Q^2\in[-m_\rho^2,10]\,$GeV$^2$ is provided by
\begin{equation}
G^{01}(Q^2) \stackrel{\rm interpolation}{=}
\frac{1+0.10\,Q^2}{1+1.073\,Q^2}\,.
\end{equation}

\subsection{Current conservation}
In Secs.\,\ref{sec:D1}--\ref{sec:D3} we have expressed formulae in terms of the baryon's unit-normalised Faddeev amplitude.  In analogy with mesons, the canonical normalisation condition amounts to an overall multiplicative rescaling so that $F_{1B}(Q^2=0)=1$ for the charged state \cite{Oettel:1999gc}.

Ward-Takahashi identities play an important role in computing the rescaling factor.  To explain, consider the contribution to $F_{1B}(Q^2)$ from Eq.\,(\ref{CalI1s}), defined as $e_u s(P)^2 F_{1B,{\cal I}^1_{s}}$, and that from Eq.\,(\ref{CalI2s}), $e_{[ud]} s(P)^2 F_{1B,{\cal I}^2_{s}}$.  Then so long as a translationally invariant regularisation scheme is used, one can show
\begin{equation}
\label{F1B0IS}
F_{1B,{\cal I}^1_{s}}(Q^2=0) = F_{1B,{\cal I}^2_{s}}(Q^2=0)\,.
\end{equation}
In addition, with definitions clear by analogy, one has
\begin{equation}
\label{F1B0IA}
F_{1B,{\cal I}^{1p}_{j}}(Q^2=0) = F_{1B,{\cal I}^{2p}_{j}}(Q^2=0)\,,\; j=1,2\,,\; p=+,0.
\end{equation}
Along with the fact that ${\cal I}^3_{j\mu}$ does not contribute at $F_{1B}(Q^2)$, then Eqs.\,(\ref{F1B0IS}), (\ref{F1B0IA}) ensure: simple additivity of the quark and diquark electric charges, and thereby guarantee a unit-charge isospin=$(+1/2)$ baryon through a single rescaling factor; and a neutral isospin=$(-1/2)$ baryon without fine tuning.  In applying our regularisation scheme, we consistently enforce Eqs.\,(\ref{F1B0IS}), (\ref{F1B0IA}).

\subsection{Typical contribution}
%
There are many terms in the complete expression for the baryon elastic electromagnetic form factors: according to one enumeration scheme, eleven each for $F_{1B}$ and $F_{2B}$.  Hence, we choose only to list one pair as an example; namely, that determined from Eq.\,(\ref{CalI2s}).  The procedure is the same in all cases.

Using Eq.\,(\ref{CalJB}), one constructs momentum-dependent Dirac-matrices that, under a trace operation, project the $F_{1B}$ and $F_{2B}$ components of each diagram.
All of the scalar expressions thus obtained are simplified by using the kinematic conditions ($K_{0^+ (i,f)}=-\ell + P_{(i,f)}$)
\begin{subequations}
\label{kinematics1}
\begin{eqnarray}
P_f^2 & = & -m_B^2 = P_i^2,\\
P_i\cdot Q & = & -\frac{1}{2} Q^2,\\
P_i\cdot P_f & = & -m_B^2 - \frac{1}{2}Q^2,\\
K_{0^+ i}^2  &=& - m_{0^+}^2,\\
Q\cdot K_{0^+ i}  &=&  -\frac{1}{2} Q^2 ,\\
Q\cdot K_{0^+ f}  &=&  \frac{1}{2} Q^2  , \\
K_{0^+ i}\cdot K_{0^+ f} &=& - m_{0^+}^2-\frac{1}{2} Q^2.
\end{eqnarray}
\end{subequations}

A Feynman parametrisation is then employed in order to produce a single denominator from the product of three propagators which appears, and the momentum-integration variable is subsequently shifted, in our case:
\begin{equation}
\ell \to l + \alpha(P_i + \beta Q).
\end{equation}
This produces a simple denominator:
\begin{equation}
[l^2 + \omega(M^2,m_{0^+}^2,Q^2,\alpha,\beta)]^3,
\end{equation}
where $M$ is the dressed-quark mass and
\begin{eqnarray}
\nonumber
\lefteqn{\omega(\alpha,\beta,M,m_{0^+},m_B,Q^2) = M^2 (1-\alpha)}\rule{1em}{0ex} \\
&+&  \alpha (m_{0^+}^2-(1-\alpha) m_B^2  + \alpha^2 \beta (1-\beta) Q^2;
\end{eqnarray}
and a numerator that is simplified using Eqs.\,\eqref{kinematics1}, their corollaries,
\begin{subequations}
\label{kinematics2}
\begin{eqnarray}
P_i \cdot K_{0+ i} & = & -l\cdot P_i -(1-\alpha) m_B^2 - \frac{1}{2} \alpha\beta Q^2,\rule{1em}{0ex}\\
\nonumber
P_i \cdot K_{0+ f} & = & -l\cdot P_i -(1-\alpha) m_B^2 \\
&& - \frac{1}{2} (1+\alpha\beta) Q^2,\\
\ell \cdot P_i & = & l\cdot P_i-\alpha m_B^2 -\frac{1}{2} \alpha\beta Q^2,\\
\nonumber \ell \cdot K_{0+ i} & = &  -l^2 + (1-2\alpha) l\cdot P_i  - 2 \alpha\beta l\cdot Q  \\
\nonumber &&-\alpha (1-\alpha) m_B^2 \\
&& - \frac{1}{2} \alpha \beta [1+2 \alpha(1-\beta)]Q^2,\\
\ell \cdot Q &= & l\cdot P_i -\frac{1}{2}\alpha (1-2\beta) Q^2,\\
\ell \cdot K_{0+ f} & = & \ell \cdot K_{0+ i} + \ell\cdot Q\,,
\end{eqnarray}
\end{subequations}
and subsequently $O(4)$-invariance.

Finally, the momentum integral is regularised to yield
\begin{eqnarray}
\nonumber
\lefteqn{F_{1B,{\cal I}^2_s}(Q^2) = \int_0^1\! d\alpha \, d\beta\, 2 \alpha \,
[ Q^2 + 4 m_B^2 ]^{-1} }\\
\nonumber
&& \times \bigg\{\bigg[ 2 m_B (\alpha m_B + M) (4 m_B^2 (1-\alpha) + (1-2\alpha \beta) Q^2) \\
\nonumber
&&  - \frac{1}{2} \alpha^2\beta(1-2\beta) Q^4 \bigg]\bar{\cal C}_2^{\rm iu}(\omega)\,,\\
&& + \frac{1}{4}[3 Q^2 + 8 m_B^2] [\bar{\cal C}_1^{\rm iu}(\omega) - \omega\bar{\cal C}_2^{\rm iu}(\omega)]\bigg\},
\label{FinalF1BI2s}\\
\nonumber
\lefteqn{F_{2B,{\cal I}^2_s}(Q^2) = -\int_0^1\! d\alpha \, d\beta\, 2 \alpha \,
2 m_B [ Q^2 + 4 m_B^2 ]^{-1}  }\\
\nonumber
&& \times \bigg\{\bigg[4 m_B^2(\alpha m_B +M)(1-\alpha) + [ (1-2\alpha\beta)M \\
\nonumber
&& + \alpha (1-\alpha\beta[1+2\beta])m_B ]Q^2\bigg\}\bar{\cal C}_2^{\rm iu}(\omega)\,,\\
&& - m_B^2 [\bar{\cal C}_1^{\rm iu}(\omega) - \omega\bar{\cal C}_2^{\rm iu}(\omega)]\bigg\},
\end{eqnarray}
where
\begin{equation}
{\cal C}_2^{\rm iu}(\omega) = (\omega^2/2) {\cal C}^{{\rm iu}\,\prime\prime}(\omega)= \frac{\omega}{2} \left({\rm e}^{-\omega \tau_{\rm uv}^2} - {\rm e}^{-\omega \tau_{\rm ir}^2}\right),\label{FinalF2BI2s}
\end{equation}
$\overline{\cal C}_2^{\rm iu}(\omega)={\cal C}_2^{\rm iu}(\omega)/\omega^2$, is a derived form of Eq.\,\eqref{RegC}.

In computing the Roper elastic form factor there is one modification at this point, arising in connection with the zero we have inserted in the associated Faddeev equation (see the last paragraph of App.\,\ref{sec:Faddeev}).  Namely, the functions ${\cal C}^{\rm iu}$ are replaced by functions
\begin{equation}
{\cal R}^{\rm iu} = {\cal C}^{\rm iu} - 2 d_{\cal F} {\cal D}^{\rm iu} + d_{\cal F}^2 {\cal H}^{\rm iu}
\end{equation}
where
\begin{eqnarray}
\nonumber
\lefteqn{{\cal H}^{\rm iu}(\omega(M^2,\alpha,P^2))= \int_0^\infty ds\,s\,\frac{s^2}{s+\omega}}\\
&\to&  \int_{\tau_{\rm uv}^2}^{\tau_{\rm iu}^2} d\tau\, \frac{6}{\tau^4} \,
\exp\left[-\tau \omega(M^2,\alpha,P^2)\right],
\end{eqnarray}
${\cal H}^{\rm iu}_1(z) = - z (d/dz){\cal H}^{\rm iu}(z)$, $\overline{\cal H}_1(z) = {\cal H}_1(z)/z$ and ${\cal H}_2^{\rm iu}(z) = (z^2/2) {\cal H}^{{\rm iu}\,\prime\prime}(z)$, $\bar {\cal H}_2^{\rm iu}(z)/z^2$.
Through this expedient we represent the square of a Faddeev amplitude that possesses a zero, as would appear in computing the elastic form factor of the excitation.

\subsection{Dressed-quark anomalous magnetic moment}
\label{App:AMM}
In the presence of dynamical chiral symmetry breaking, a dressed light-quark possesses a large anomalous electromagnetic moment \cite{Chang:2010hb}.  To indicate the effect on form factors that one might expect from this phenomenon, we modified the quark-photon coupling as follows:
\begin{equation}
\label{DQAMM}
\Gamma(Q) = \gamma_\mu^{T} P_T(Q^2) + \frac{\zeta}{2 M} \sigma_{\mu\nu} Q_\nu \exp(-Q^2/4 M^2)
\end{equation}
where $M$ is the dressed-quark mass.  Both the value of $\zeta=1/2$ and the rate at which the anomalous moment term decays are taken from the distribution computed in Ref.\,\cite{Chang:2010hb}.

The anomalous moment has no effect on the elastic form factor of the scalar diquark but it does change the form factors of the axial-vector diquarks; viz., with our standard parameter choice, Table~\ref{Table:static}, and $z=Q^2$, the following functions provide an accurate interpolation of the result:
\begin{subequations}
\begin{eqnarray}
F_1^{1^+}(z) & = & \frac{(1 + 0.98 z)}{1 + 2.75 z + 1.26 z^2}, \\
F_2^{1^+}(z) & = & -\frac{3.23 + 0.048 z}{1 + 2.11 z + 0.0037 z^2},\\
F_3^{1^+}(z) & = & \frac{1.19 + 0.33 z}{1 + 1.38 x + 4.62 z^2}.
\end{eqnarray}
\end{subequations}
Comparison with Eqs.\,\eqref{GsFs}, \eqref{GsAVqq} reveals that the dressed-quark anomalous electromagnetic moment in Eq.\,\eqref{DQAMM} increases the axial-vector diquarks' magnetic moment by 50\% and the magnitude of its quadrupole moment by 30\%.

\section{Transition Current}
\label{App:transitioncurrent}
With the baryon spinors we have defined, the current in Eq.\,\ref{NRtransition} can be expressed
\begin{eqnarray}
\nonumber
J_\mu^\ast(P_f,P_i) & = & i e\,  \Lambda^R_+(P_f)\left[ \gamma_\mu^T F_{1\ast}(Q^2) \right. \\
&& \left. + \frac{1}{M_R+M_{N}} \sigma_{\mu\nu} Q_\nu F_{2\ast}(Q^2)\right] \Lambda^N_+(P_i) , \rule{2em}{0ex}
\label{ANRtransition}
\end{eqnarray}
where the positive-energy projection operators are as defined in Eq.\,(\ref{Lplus}).  The same three diagrams contribute to the transition but with the modification that the final state is the Roper resonance.  This means that the kinematics are different, Eq.\,\eqref{transitionkinematics}, and in Eq.\,\eqref{CalI1s}, for example,
\begin{equation}
s(P)^2 \to s_R(P_f) s_N(P_i)\,.
\end{equation}
With such changes implemented throughout, the analysis proceeds unchanged, although one must pay attention to the modified kinematics when computing invariants, Eqs.\,\eqref{kinematics1}, and working through the Feynman parametrisation, Eqs.\,\eqref{kinematics2}, until final expressions, such as those in Eqs.\,\eqref{FinalF1BI2s}, \eqref{FinalF2BI2s}, are obtained.

At this point, the functions ${\cal C}^{\rm iu}$ are replaced by the functions ${\cal F}^{\rm iu}$, in order that the zero we have inserted into the Roper's Faddeev amplitude is expressed in the transition form factors.

We require that the Roper's dressed-quark core be orthogonal to that of the nucleon and insist that each radially excited state possess a zero in its Faddeev amplitude, as in Ref.\,\cite{Roberts:2011cf}.  The latter requirement ensures that the contact interaction is able to produce a radial excitation of both the $\Delta$ resonance and its parity partner.  On the other hand, it modifies the Faddeev kernel, so that the nucleon kernel is different from that for the Roper and therefore orthogonality is not assured.  This drawback, which accompanies the interaction's simplicity, is readily corrected now that we have expressions for the transition form factors.

As mentioned above, orthogonality means that $F_{1*}(Q^2=0)=0$ for both the charged and neutral resonances.  (The analogue of this condition has been used in studies of meson radial excitations, both with momentum-independent \cite{Volkov:1996br,Volkov:1999yi} and momentum-dependent kernels \cite{Krassnigg:2003wy,Holl:2004fr}.)  In Ref.\,\cite{Roberts:2011cf}, lacking expressions for the transition form factors, the location of the zero in the Roper's Faddeev amplitude was fixed following inspection of its position in meson Bethe-Salpeter amplitudes.  This led to the choice  $M^2 d_{\cal F} = 1.0 \pm 0.2$.

Herein, we first consider $F_{1R^+\to p}(Q^2=0)$.  Employing Eq.\,\eqref{ANRtransition}, the analogues of Eqs.\,\eqref{CalI1s}, \eqref{CalI2s}, \eqref{calI3j}, and using Eqs.\,\eqref{isospinFF}, one finds that $F_{1R^+\to p}(Q^2=0)$ receives just one contribution; viz., that of Diagram~1 where the photon strikes a dressed-quark in association with a scalar diquark (all others are zero at $Q^2=0$).  Orthogonality of the proton and charged-Roper is then assured if
\begin{equation}
\frac{1}{d_{\cal F}} = 0.77 M^2\,,
\label{RealdF}
\end{equation}
a value just 3\% smaller than the lower bound estimated in Ref.\,\cite{Roberts:2011cf} so that the mass estimate therein ($1.82 \pm 0.07\,$GeV) was reasonable.  In fact, with Eq.\,\eqref{RealdF} one obtains the Roper mass in Table~\ref{Table:FE}B, which is in even better agreement with EBAC's result for the dressed-quark core; viz., $1.76\pm 0.1\,$GeV, Eq.\,\eqref{Ropermass}.

This procedure does not fix the value of $s_R$.  For guidance in this respect we turn again to studies of meson excitations.  At zero relative momentum in a radial-excitation's Bethe-Salpeter amplitude, the magnitude of the dominant Dirac stucture's leading Chebyshev moment is approximately one-half of that for the ground-state \cite{Holl:2004fr,Qin:2011xq}.  We therefore choose
\begin{equation}
s_R = -\frac{1}{2} s_N = -0.44\,,
\end{equation}
as listed in Table~\ref{Table:FE}B.  The sign here matches that produced by the Roper's Faddeev equation but the magnitude is five-times larger: the Faddeev equation for the Roper produces a state that is 99\% axial-vector diquark.

Now, given the canonical normalisation condition, $F_{1R^+}(Q^2=0)=1$, and Eqs.\,\eqref{aia0}, there is only one entry left to be fixed in the Roper's Faddeev amplitude.  That is set by the condition $F_{1 R^0\to n}(Q^2=0)=0$, whose only nonzero entries are Diagram~1 quark plus axial-vector diquark contributions.  We thus arrive at the Faddeev amplitude entries for the Roper in Table~\ref{Table:FE}B.

\bigskip

\section{Euclidean Conventions}
\label{App:EM}
In our Euclidean formulation:
\begin{equation}
p\cdot q=\sum_{i=1}^4 p_i q_i\,;
\end{equation}
\begin{eqnarray}
&& \{\gamma_\mu,\gamma_\nu\}=2\,\delta_{\mu\nu}\,;\;
\gamma_\mu^\dagger = \gamma_\mu\,;\;
\sigma_{\mu\nu}= \frac{i}{2}[\gamma_\mu,\gamma_\nu]\,; \rule{2em}{0ex}\\
&& {\rm tr}\,[\gamma_5\gamma_\mu\gamma_\nu\gamma_\rho\gamma_\sigma] =
-4\,\epsilon_{\mu\nu\rho\sigma}\,, \epsilon_{1234}= 1\,.
\end{eqnarray}

A positive energy spinor satisfies
\begin{equation}
\label{DiracEqn}
\bar u(P,s)\, (i \gamma\cdot P + M) = 0 = (i\gamma\cdot P + M)\, u(P,s)\,,
\end{equation}
where $s=\pm \frac{1}{2}$ is the spin label.  The spinor is normalised:
\begin{equation}
\bar u(P,s) \, u(P,s) = 2 M \,,
\end{equation}
and may be expressed explicitly:
\begin{equation}
u(P,s) = \sqrt{M- i {\cal E}}
\left(
\begin{array}{l}
\chi_s\\
\displaystyle \frac{\vec{\sigma}\cdot \vec{P}}{M - i {\cal E}} \chi_s
\end{array}
\right)\,,
\end{equation}
with ${\cal E} = i \sqrt{\vec{P}^2 + M^2}$,
\begin{equation}
\chi_+ = \left( \begin{array}{c} 1 \\ 0  \end{array}\right)\,,\;
\chi_- = \left( \begin{array}{c} 0\\ 1  \end{array}\right)\,.
\end{equation}
For the free-particle spinor, $\bar u(P,s)= u(P,s)^\dagger \gamma_4$.

The spinor can be used to construct a positive energy projection operator:
\begin{equation}
\label{Lplus} \Lambda_+(P):= \frac{1}{2 M}\,\sum_{s=\pm} \, u(P,s) \, \bar
u(P,s) = \frac{1}{2M} \left( -i \gamma\cdot P + M\right).
\end{equation}


A charge-conjugated Bethe-Salpeter amplitude is obtained via
\begin{equation}
\label{chargec}
\bar\Gamma(k;P) = C^\dagger \, \Gamma(-k;P)^{\rm T}\,C\,,
\end{equation}
where ``T'' denotes a transposing of all matrix indices and
$C=\gamma_2\gamma_4$ is the charge conjugation matrix, $C^\dagger=-C$.  We note that
\begin{equation}
C^\dagger \gamma_\mu^{\rm T} \, C = - \gamma_\mu\,, \; [C,\gamma_5] = 0\,.
\end{equation}


\end{document}